\documentclass[prd,twocolumn,superscriptaddress,nofootinbib,amsmath,amssymb,preprintnumbers,floatfix]{revtex4-1}
\usepackage{graphicx,epsfig,psfrag,bm,amssymb}
\usepackage{dcolumn}
\usepackage{hyperref}
\usepackage{bm}
\usepackage{color}
\usepackage{mathrsfs,amsfonts,hepunits, color}
\usepackage[utf8]{inputenc}
\usepackage{xcolor}
\usepackage{braket}

\newcommand{\be}{\begin{equation}}
\newcommand{\ee}{\end{equation}}

\newcommand{\RCM}{\vec{R}_{\rm CM}}

\newcommand{\azero}{{\rm a}_0}

\begin{document}

\title{The Molecular Migdal Effect}

\author{Carlos Blanco}
\email{carlosblanco2718@princeton.edu}
\affiliation{Stockholm University and The Oskar Klein Centre for Cosmoparticle Physics,  Alba Nova, 10691 Stockholm, Sweden}
\affiliation{Department of Physics, Princeton University, Princeton, New Jersey 08544, U.S.A.}

\author{Ian Harris}
\email{ianwh2@illinois.edu}
\affiliation{Department of Physics, University of Illinois Urbana-Champaign, Urbana, Illinois 61801, U.S.A.}

\author{Yonatan Kahn}
\email{yfkahn@illinois.edu}
\affiliation{Department of Physics, University of Illinois Urbana-Champaign, Urbana, Illinois 61801, U.S.A.}
\affiliation{Illinois Center for Advanced Studies of the Universe, University of Illinois Urbana-Champaign, Urbana, Illinois 61801, U.S.A.}

\author{Benjamin Lillard}
\email{blillard@illinois.edu}
\affiliation{Department of Physics, University of Illinois Urbana-Champaign, Urbana, Illinois 61801, U.S.A.}
\affiliation{Illinois Center for Advanced Studies of the Universe, University of Illinois Urbana-Champaign, Urbana, Illinois 61801, U.S.A.}

\author{Jes\'{u}s P\'{e}rez-R\'{i}os}
\email{jesus.perezrios@stonybrook.edu}
\affiliation{Department of Physics and Astronomy, Stony Brook University, Stony Brook, New York  11794, U.S.A.}
\affiliation{Institute for Advanced Computational Science, Stony Brook University, Stony Brook, New York 11794, U.S.A.}

\begin{abstract}
Nuclear scattering events with large momentum transfer in atomic, molecular, or solid-state systems may result in electronic excitations. In the context of atomic scattering by dark matter (DM), this is known as the Migdal effect, but the same effect has also been studied in molecules in the chemistry and neutron scattering literature. Here we present two distinct Migdal-like effects from DM scattering in molecules, which we collectively refer to as the molecular Migdal effect: a center-of-mass recoil, equivalent to the standard Migdal treatment, and a non-adiabatic coupling resulting from corrections to the Born-Oppenheimer approximation. The molecular bonds break spherical symmetry, leading to large daily modulation in the Migdal rate from anisotropies in the matrix elements. Our treatment reduces to the standard Migdal effect in atomic systems but does not rely on the impulse approximation or any semiclassical treatments of nuclear motion, and as such may be extended to models where DM scatters through a long-range force. We demonstrate all of these features in a few simple toy models of diatomic molecules, namely ${\rm H}_2^+$, N$_2$, and CO, and find total molecular Migdal rates competitive with those in semiconductors for the same target mass. We discuss how our results may be extended to more realistic targets comprised of larger molecules which could be deployed at the kilogram scale.

\end{abstract}

\maketitle

\section{Introduction}

The Migdal effect, in which nuclear scattering leads to a visible electron recoil, is a promising avenue to detect sub-GeV dark matter (DM) scattering with nuclei. Such light DM is kinematically mismatched with nuclei and thus leads to very small elastic scattering energies, often below detection thresholds. However, because electrons and nuclei are coupled in all atomic, molecular, and solid-state systems, perturbations to the nuclei can induce electronic transitions. The probability of this transition is small, but since electronic transition energies are at the eV scale which is above the thresholds of existing detectors, even a small rate is favorable compared to an unobservable elastic scattering signal. The Migdal effect in atoms, in which recoiling nuclei lead to atomic excitation or ionization, has a long and interesting history, first proposed nearly a century ago in the context of alpha and beta decay~\cite{migdal1941ionization} and subsequently rediscovered by the WIMP DM community~\cite{Bernabei:2007jz}. Independently, the neutron scattering community invoked nucleus-electron correlations similar to the atomic Migdal effect to explain anomalous cross sections in compounds containing hydrogen~\cite{gidopoulos2005breakdown,reiter2005origin,colognesi2005can}. The Migdal effect in the context of DM has been calculated for isolated atoms~\cite{Ibe:2017yqa,Liu:2020pat} and semiconductors~\cite{Knapen:2020aky,Liang:2020ryg,Liang:2022xbu} (see also~\cite{Dolan:2017xbu,Bell:2019egg,Baxter:2019pnz,Essig:2019xkx,Liang:2019nnx,GrillidiCortona:2020owp,Wang:2021oha} for additional theoretical investigations of the Migdal effect), and there is an active program to try to measure the ionizing Migdal effect experimentally using Standard Model probes~\cite{Nakamura:2020kex,Bell:2021ihi,Araujo:2022wjh}. 

In this paper, we present for the first time two distinct directional Migdal-like effects in \emph{excitation} of molecules that we call collectively the molecular Migdal effect. This can be seen as the low-energy complement of the \emph{ionizing} Migdal effect of core electron shells of atoms bound in molecules (see e.g. Ref.~\cite{Araujo:2022wjh}), which is isotropic and does not depend significantly on the molecular nature of the nuclear or electronic states. We focus specifically on diatomic molecules, treating them as toy examples useful in order to derive analytic expressions for the matrix elements, identify the origin of anisotropy and directionality for the molecular Migdal effect, and determine parametric scalings which can generalize to larger molecules. Due to the anisotropy inherent in the molecular states, we predict order-1 daily modulation of the Migdal signal for DM masses of 10 MeV to 1 GeV. Our qualitative results should generalize to well-oriented molecules with weak intermolecular interactions, such as aromatic organic compounds that can form molecular crystals and which already serve as excellent anisotropic targets for DM-electron scattering which could conceivably be deployed at the kilogram scale~\cite{Blanco:2019lrf,Blanco:2021hlm}. Therefore, we will discuss the path to extending our formalism to larger molecules and how we may use existing molecular data to identify potential targets with large molecular Migdal rates. 

We base our treatment largely on a series of papers formulating the cross section for molecular excitations following neutron scattering~\cite{lovesey1982electron,elliott1984study,gidopoulos2005breakdown,reiter2005origin,colognesi2005can}. We rederive and adapt for sub-GeV DM scattering the following results: 
\begin{itemize}
\item Migdal excitation has a component proportional to an electronic dipole matrix element, $\langle \psi_f | \vec{r} | \psi_i \rangle$, where $|\psi_i \rangle$ and $|\psi_f \rangle$ are the initial and final electronic states, respectively. In previous work this was understood as arising from a semiclassical approximation for the struck nucleus for a contact interaction, but here we show that it arises simply from the mismatch between the center of mass (COM) of the nuclei and the COM of the entire molecule including the electrons. This component of the Migdal effect, which we refer to as the center-of-mass recoil (CMR), thus requires no restrictions on the size of the momentum transfer and holds equally well for scattering through a long-range force. \footnote{Despite the nomenclature, the CMR effect exists even if the COM is fixed and does not actually recoil.}
\item There is a second component of the Migdal excitation probability, arising from corrections to the Born-Oppenheimer (BO) approximation. Such an effect does not exist for atoms with a single nucleus, but instead describes the behavior of molecular systems where electronic and nuclear motion may be parametrically separated because of the small ratio $m_e/M$, where $M$ is the nuclear mass. The many-body ground state of the molecule contains admixtures of excited electronic states with coefficients of order $m_e/M$, referred to in the literature as a non-adiabatic coupling (NAC), a nomenclature we adopt. The NAC gives the ground state a nonzero overlap with excited electronic states, yielding a Migdal matrix element proportional to $\langle \psi_f | \nabla_\rho \psi_i \rangle$ where  $\nabla_\rho$ is the gradient with respect to the nuclear separation $\vec{\rho}$.
\end{itemize}
We show that both CMR and NAC probabilities have identical parametric scalings, and compute the relevant electronic matrix elements for some simple diatomic molecules (${\rm H}_2^+$, N$_2$, and CO, selected due to their computational feasibility) to show the interplay of these two effects. Interestingly, the CMR and NAC Migdal effects have orthogonal selection rules: the dipole matrix element connects electronic states of opposite parity, while the NAC matrix element connects states of like parity.\footnote{We will focus exclusively on molecular orbitals consisting of valence electrons, but since molecular spectroscopic notation is possibly unfamiliar to some physicists, the like-parity transitions are analogous to $1s \to 2s$ in atomic hydrogen, and those of opposite parity are analogous to $1s \to 2p$.} This allows for the two molecular Migdal components to be distinguished experimentally, since these transitions typically have well-separated energies.

In addition, we point out the following new result concerning the directional dependence of the scattering. Consider a situation where the internuclear axis is fixed along a particular direction $\hat{\rho}_0$. This is perhaps unrealistic for diatomic molecules, but accurately describes a molecular crystal where molecules have a fixed orientation within a unit cell because the crystal spontaneously breaks rotational invariance. Defining an anisotropy parameter
\be
\label{eq:etadef}
\eta = \hat{q} \cdot \hat{\rho}_0,
\ee
where $\vec{q}$ is the momentum transfer of the interaction, there are now two sources of anisotropy in the Migdal excitation probability:
\begin{itemize}
\item Both CMR and NAC matrix elements inherit the anisotropy of the electronic wavefunctions, since the dipole matrix element and the NAC gradient matrix element both point along the direction of the molecular axis $\hat{\rho}_0$. In both cases the Migdal probability carries a factor of $\eta^2$.
\item Both CMR and NAC contain nuclear matrix elements schematically of the form $\langle \chi_f | e^{i \vec{q} \cdot \vec{\rho}} | \chi_i \rangle$ where $|\chi_i \rangle$ and $| \chi_f \rangle$ are nuclear states. Squaring and evaluating this matrix element yields additional anisotropy of the form $\eta^{2n} \exp \left( -\frac{q^2}{2 \mu \omega} \eta^2 \right)$, where $\mu$ is the reduced mass of the nuclei, $\omega$ is a characteristic vibrational frequency, and $n$ depends on the vibrational final state with the largest overlap with the initial state. The factor $q^2/(2 \mu \omega)$ in the exponential can be order-1 for sub-GeV DM and thus the directionality of the scattering rate depends strongly on the DM mass. 
\end{itemize}
The anisotropy of the Migdal excitation probability leads to the appealing possibility of directional detection, which (as in the case of DM-electron scattering) does not depend on observing the direction of any final states, but rather yields a sidereal daily modulation in the rate of e.g.\ photons emitted from the de-excitation of the excited molecular state. While diatomic molecules have already been proposed as possible targets for DM-nuclear scattering~\cite{Essig:2016crl,Essig:2019kfe}, in a typical experimental situation with gas detectors, the molecules will be in their rotational ground state, which is isotropic. Thus, the directional dependence we have identified will average out and disappear. However, in larger molecules with fixed orientation, for example organic scintillator crystals, the large daily modulation should persist. As we will show, the daily modulation from the molecular Migdal effect is not a threshold effect, and persists at the $\mathcal{O}(1)$ level even for DM masses well above the kinematic threshold for electronic excitation.

This paper is organized as follows. In Sec.~\ref{sec:BO}, we review the non-adiabatic corrections to the BO approximation in diatomic molecules, compute the electronic transition probability following a nuclear scattering event to leading order in $m_e/M$, and identify the CMR and NAC components of the molecular Migdal effect. In Sec.~\ref{sec:Daily}, we compute the nuclear and electronic matrix elements for our three diatomic toy examples and show the daily modulation of the electronic excitation rate as a function of the DM mass, demonstrating that the NAC contribution typically dominates and gives competitive sensitivity to semiconductor targets. In Sec.~\ref{sec:LargerMolecules} we outline how our results may be extended to larger molecules. We conclude in Sec.~\ref{sec:Conclusions}.

\section{The Molecular Migdal effect}
\label{sec:BO}

\begin{figure}
    \centering
    \includegraphics[width=0.4 \textwidth]{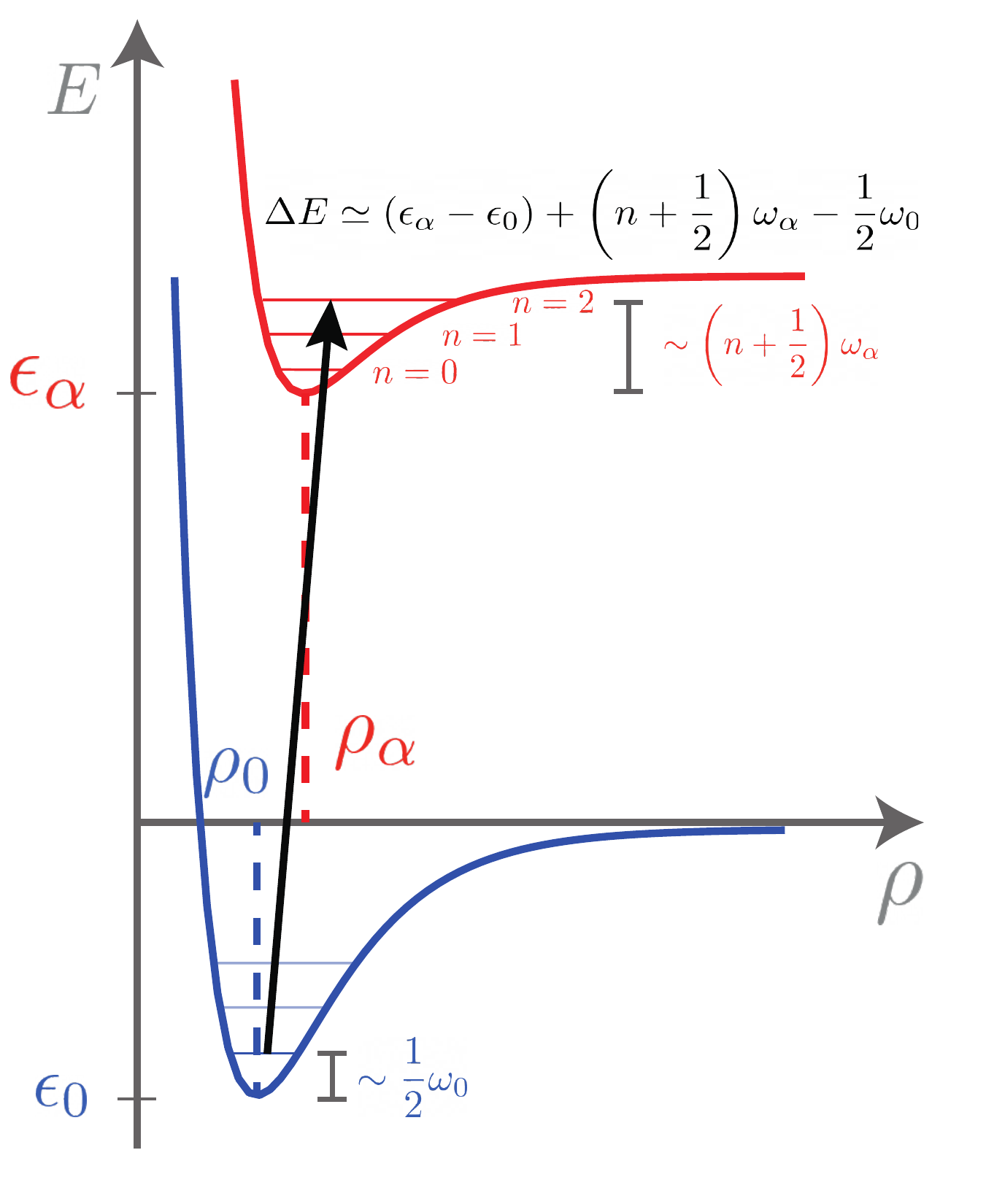}
    \vspace{-0.4cm}
    \caption{Cartoon illustrating the transition from the electronic ground state (blue) to an excited electronic state (red). The energy manifolds which goven the nuclear states are determined by the electronic configuration and are modeled by a Morse potential as a function of the internuclear separation $\rho$. Each electronic state is split by vibrational substates which are approximately harmonic near the potential minima.}
    \label{fig:TransitionDiagram}
    \vspace{-0.3cm}
\end{figure}

A diatomic molecule has a many-body wavefunction $\Psi(\vec{R}_1, \vec{R}_2, \vec{r}_{e,i})$ where $\vec{R}_{1,2}$ are the nuclear positions, $\vec{r}_{e,i}$ are the electron positions, and $i$ runs over all of the electrons in the molecule. In the BO approximation, the electrons are treated as responding instantaneously to changes in the nuclear positions, and the wavefunction factorizes into a product of nuclear and electronic wavefunctions. This factorization only holds in the strict limit $m_e/M_i \to 0$, though, where $M_i$ are the nuclear masses. To see this, we start from the Schr\"{o}dinger equation for the molecule:
\be
\frac{1}{m_e} \sum_i \nabla_i^2 \Psi + \left(\frac{\nabla_{R_1}^2 }{M_1}+ \frac{ \nabla_{R_2}^2}{M_2}\right)\Psi + 2(E - V) \Psi = 0
\label{eq:SEall}
\ee
where $E$ is the energy of the state $\Psi$, and $V = V_{ee} + V_{eN} + V_{NN}$ is the potential which contains electron-electron, electron-nucleus, and nucleus-nucleus interactions, respectively. The BO ansatz is
\be
\Psi(\vec{R}_1, \vec{R}_2, \vec{r}_{e,i}) \approx \theta(\RCM)\chi^{(\alpha)}(\vec{R}_1, \vec{R}_2) \psi_\alpha(\vec{r}_i; \rho_\alpha).
\label{eq:BE}
\ee
Here, $\theta$ is the overall COM motion, $\chi^{(\alpha)}$ depends only on the nuclear positions, and $\psi_\alpha$ is an electronic wavefunction which depends parametrically on the equilibrium separation between the two nuclei, $\rho_\alpha$, and for which the electronic coordinate $\vec{r}_i$ is taken with respect to the COM of the molecule, $\vec{r}_i = \vec{r}_{e,i} - \RCM$. 

Neglecting the COM motion which always factors out, the Schr\"{o}dinger equation approximately separates as
\begin{align}
    & \frac{1}{m_e} \sum_i \nabla_i^2 \psi_\alpha + 2(\epsilon_\alpha - V_{ee} - V_{eN}) \psi_\alpha = 0, \label{eq:ElectronicEqSep}\\
    & \left(\frac{\nabla_{R_1}^2}{M_1} + \frac{\nabla_{R_2}^2}{M_2} \right)\chi^{(\alpha)} + 2(E - \epsilon_\alpha - V_{NN})\chi^{(\alpha)} = 0.
\end{align}
The first equation determines the electronic state $\psi_\alpha$ and its energy eigenvalue $\epsilon_\alpha$ for fixed nuclear positions, and the second determines the nuclear state given $\epsilon_\alpha$ as a function of nuclear positions. Note that the equilibrium separation $\rho_\alpha$ is determined by minimizing the effective potential governed by $\epsilon_\alpha$, and thus depends on the electronic state $\alpha$, as illustrated in Fig.~\ref{fig:TransitionDiagram}. Likewise, vibrational excitations above this equilibrium state depend on $\alpha$, which we emphasize with our notation $\chi^{(\alpha)}$.

The terms neglected in the separation of the Schr\"{o}dinger equation are of the form $\frac{1}{M_{1,2}} \chi^{(\alpha)} \nabla^2_{1,2} \psi_\alpha$ and $\frac{2}{M_{1,2}} (\nabla_{1,2} \psi_\alpha) (\nabla_{1,2} \chi^{(\alpha)})$. As anticipated, these vanish as $M_{1,2} \to \infty$, but treated as a perturbation to $V_{eN}$ in time-independent perturbation theory, they will correct the electronic wavefunctions, leading to the NAC Migdal effect \cite{lovesey1982electron} as we describe further in Sec.~\ref{sec:NAC} below.

The relevant squared matrix element for a Migdal transition to a particular electronic state $\psi_\alpha$ in a diatomic molecule, through a momentum deposit $\vec{q}$ from the DM, is
\be
\label{eq:PM}
P^{(\alpha)} = \sum_{'} | \langle \Psi'_\alpha | a_1 e^{i \vec{q} \cdot \vec{R}_1} + a_2 e^{i \vec{q} \cdot \vec{R}_2} | \Psi_0 \rangle |^2
\ee
where $\Psi_0$ is the molecular ground state and the sum over $\Psi'_\alpha$ contains all final nuclear states $\chi^{(\alpha)}$ associated with $\psi_\alpha$.\footnote{Strictly speaking, Eq.~(\ref{eq:PM}) should contain an energy-conserving delta function $\delta(E' - E_0)$ inside the sum, but since the nuclear energies are much smaller than the electronic energies $\epsilon_\alpha$, $E' \approx \epsilon_\alpha$ for all terms in the sum and the delta function can be approximately factored out. We will restore the delta function in Sec.~\ref{sec:Daily} below.} We have allowed for the possibility that DM may couple differently to nuclei 1 and 2 by including arbitrary (real) coefficients $a_1$ and $a_2$ (the analogues of different neutron scattering lengths in the case of neutron-molecule scattering). For sub-GeV DM, the momentum transfer is always smaller than the inverse nuclear radius, so the interaction is always coherent over the nucleus and the nuclear form factor is unity. A long-range DM interaction may be accommodated by adding a factor of $1/q^2$ in the matrix element, as well as screening effects by e.g.\ multiplying $a_{1,2}$ by atomic form factors. 

Because we have in mind the application of the Migdal effect to solid-state systems, in particular scintillation transitions in molecular crystals, we will narrow our focus from the general expression~(\ref{eq:PM}) in two ways:
\begin{enumerate}
    \item We will only consider \emph{bound} final states, both for the electrons and the nuclei, as shown in Fig.~\ref{fig:TransitionDiagram}. The total Migdal rate, which includes both ionized electron states and dissociated nuclear states, will necessarily be larger, but the signals are expected to be experimentally distinct. The signature of a single electronic excitation is a narrow spectral line. In contrast, ionization of inner shell electrons leads to broad, energetic spectra that must be distinguished from the ionization accompanying the elastic nuclear recoil~\cite{Araujo:2022wjh}. 
    \item We will neglect both COM motion of the molecule and rotational excitations, since these will be highly suppressed in a crystal compared to vibrational modes. In particular, we do not take the nuclear ground state to be the isotropic rotational ground state where the direction of the molecular axis is undetermined, but rather fix $\hat{\rho}_0 = \hat{z}$, and likewise for the excited nuclear states.\footnote{Our isolation of vibrational nuclear states from the rotational motion of the molecule may be seen as focusing on the normal modes of the molecule, with diatomic molecules having only a single normal mode, but with polyatomic molecules hosting many more.} In this setup, the nuclear wavefunctions $\chi_n^{(\alpha)}(\rho)$ are then a function only of the nuclear separation,
    \be
    \label{eq:rhodef}
    \rho = |\vec{R}_2 - \vec{R}_1|,
    \ee
    and may be labeled by a single integer $n$ characterizing the vibrational level. Furthermore, all dot products of the form $\vec{q} \cdot \vec{\rho}$ can then be written as $q \rho \eta$, where $\eta$ is the anisotropy parameter defined in Eq.~(\ref{eq:etadef}).
\end{enumerate}
In what follows, we will compute Eq.~(\ref{eq:PM}) to leading order in $m_e/M_{1,2}$, and find schematically
\begin{align}
P^{(\alpha)} & = P^{(\alpha)}_{\rm CMR} + P^{(\alpha)}_{\rm NAC} \\
& = P^{(\alpha)}_N \times \left(P^{(\alpha)}_{e, \rm{CMR}} + P^{(\alpha)}_{e, \rm{NAC}}\right) \\
& \sim \mathcal{O}(1) \times \left(\frac{m_e}{M}\right)^2 (q \azero)^2
\label{eq:PalphaSchematic}
\end{align}
where $\azero$ is the Bohr radius, $M = M_1 + M_2$, and $P^{(\alpha)}_N$ and $P^{(\alpha)}_e$ are squared nuclear and electronic matrix elements, respectively. In particular we will find that the nuclear matrix elements for both CMR and NAC are order-1 for states $\alpha$ with large nuclear wavefunction overlaps with the ground state, and that the CMR and NAC electronic matrix elements have identical parametric scalings as shown in Eq.~(\ref{eq:PalphaSchematic}).

\subsection{CMR Migdal Effect}
\label{sec:CMR}

In a diatomic molecule, the individual nuclear coordinates $\vec{R}_{1,2}$ are related to the COM and relative coordinates as follows:
\begin{align}
\vec{R}_1 & = \RCM - \frac{\mu}{M_1} \vec{\rho} - \frac{m_e}{M}\sum_{i}\vec{r}_i \label{eq:R1} \\
\vec{R}_2 & = \RCM + \frac{\mu}{M_2} \vec{\rho} - \frac{m_e}{M}\sum_{i} \vec{r}_i
\label{eq:R2}
\end{align}
where $\mu = M_1 M_2/M$ is the reduced nuclear mass and $\vec{\rho} = \vec{R}_2 - \vec{R}_1$ is the nuclear separation vector.

Note that because the COM of the molecule includes contributions from the electronic coordinates, the nuclear coordinates contain admixtures of the relative electron coordinates with coefficients $m_e/M$. As a result, the CMR contribution to $P^{(\alpha)}$ is
\be
P^{(\alpha)}_{\rm CMR} = |\langle \psi_\alpha | e^{-i \frac{m_e}{M} \vec{q} \cdot \sum_{i}\vec{r}_i} | \psi_0 \rangle|^2 \times P^{(\alpha)}_{N,{\rm CMR}}
\ee
where $P^{(\alpha)}_{N,{\rm CMR}}$ is the squared nuclear matrix element
\be
\label{eq:PNCMR}
P^{(\alpha)}_{N,{\rm CMR}} = \sum_{n} | \langle \chi^{(\alpha)}_n | a_1 e^{-i \frac{\mu}{M_1}q \rho \eta} + a_2 e^{+i \frac{\mu}{M_2} q \rho \eta} | \chi_0 \rangle|^2
\ee
summed over vibrational states $\chi^{(\alpha)}_n$ associated with the electronic state $\alpha$. In the particular case of a homonuclear diatomic molecule, where $a_1 = a_2 \equiv a$ and $\mu/M_1 = \mu/M_2 = 1/2$, we have
\be
P^{(\alpha)}_{N,{\rm CMR}} \to 4 a^2  \sum_{n} | \langle \chi^{(\alpha)}_n | \cos (q \rho \eta/2)| \chi_0 \rangle|^2 \ \  {\rm (hom.)}
\ee
In the electronic matrix element, the typical kinematics of sub-GeV DM are such that $(m_e/M)q \ll \azero$ and thus the exponential may be Taylor-expanded to yield an electronic excitation probability
\be
\label{eq:PeCMR}
P^{(\alpha)}_{e, {\rm CMR}} = \left(\frac{m_e}{M}\right)^2 \left | \, \vec{q} \cdot  \Big\langle \psi_\alpha \Big| \sum_{i}\vec{r}_i \Big| \psi_0 \Big\rangle \right |^2,
\ee
analogous to similar results for atomic systems which have been obtained under various sets of assumptions~\cite{Ibe:2017yqa,Baxter:2019pnz,Essig:2019xkx,Knapen:2020aky,Kahn:2021ttr}. For diatomic molecules, the dipole matrix element will always point along the molecular axis, and therefore for fixed orientation $\hat{\rho}_0$, we can write
\be
\langle \psi_\alpha | \sum_{i}\vec{r}_i | \psi_0 \rangle \equiv D_{\alpha 0}\, \hat{\rho}_0
\ee
and 
\be
\label{eq:PeCMRWithEta}
P^{(\alpha)}_{e, {\rm CMR}} = \left(\frac{m_e}{M}\right)^2 q^2 \eta^2 |D_{\alpha 0}|^2.
\ee

We note that this result was also derived earlier in the context of neutron scattering in Ref.~\cite{colognesi2005can}. Furthermore, $D_{\alpha 0}$ can be experimentally determined using spectroscopy since it is essentially the oscillator strength of the transition, allowing a data-driven prediction of the CMR Migdal rate~\cite{Liu:2020pat}. Since $P^{(\alpha)}_{e, {\rm CMR}}$ is already proportional to $(m_e/M)^2$, we do not need to include the non-adiabatic corrections to $\Psi_0$ or $\Psi'_\alpha$ at this order.

\subsection{NAC Migdal Effect}
\label{sec:NAC}
The NAC component of the molecular Migdal effect arises from corrections to the wavefunctions rather than the coordinates, so we may ignore the electronic coordinates in Eqs.~(\ref{eq:R1})--(\ref{eq:R2}) to leading order in $m_e^2/M^2$. We then compute the matrix element in Eq.~(\ref{eq:PM}) as follows, setting $\vec{R}_{\rm CM} = 0$ as we are ignoring COM motion:
\be
\mathcal M^{(\alpha)}_{\rm NAC}
= \langle \Psi'_\alpha | a_1 e^{-i \frac{\mu }{M_1}  \vec q \cdot \vec \rho}  + a_2 e^{i \frac{\mu }{M_2}  \vec q  \cdot \vec \rho } | \Psi_0 \rangle .
\label{eq:NACmatrixEl}
\ee

We now include non-adiabatic corrections to the wavefunctions. Consider a total wavefunction $\Psi_\alpha$ which can be expressed as $\chi^{(\alpha)} (\psi_\alpha + \delta \psi_\alpha)$, where $\psi_\alpha$ is the unperturbed electronic wavefunction in the BO approximation. As we show in Appendix~\ref{app:NAC}, the effective perturbing potential in the electronic Schr\"{o}dinger equation is given by
\begin{align}
\delta V &= -\frac{1}{\mu} \frac{\nabla_{\! \rho} \chi}{\chi} \cdot \nabla_{\! \rho}.
\label{eq:deltaV}
\end{align}
Note that when the orientation of the molecular axis is fixed, $\nabla_\rho \equiv d/d\rho$ is an ordinary derivative. We can thus apply first-order perturbation theory to the electronic wavefunctions only,
\be
\delta \psi_\alpha = \sum_{\alpha' \neq \alpha} \frac{\langle \psi_{\alpha' } | \delta V | \psi_\alpha \rangle }{\epsilon_{\alpha} - \epsilon_{\alpha'} } \psi_{\alpha'},
\label{eq:deltapsi}
\ee
which shows as long as the perturbation matrix element does not vanish, the ground state with $\alpha = 0$ contains admixtures of the excited electronic states, and vice versa.

Multiplying by the nuclear wavefunction $\chi^{(\alpha)}$, we identify the first non-adiabatic correction to the molecular wavefunction,
\be
    \delta \Psi_\alpha = \frac{1}{\mu} (\nabla_\rho \chi^{(\alpha)}) \cdot \sum_{\alpha' \neq \alpha}  \frac{\vec G_{\alpha' \alpha}}{\epsilon_{\alpha'} - \epsilon_\alpha}\psi_{\alpha'}(\vec{r}_i\text{;} \rho_{\alpha'}), 
\ee
where $\epsilon_\alpha$ are the energies of the electronic states $\alpha$ (the eigenvalues of the electronic equation~(\ref{eq:ElectronicEqSep}), and the non-adiabatic coupling vectors $\vec G_{\alpha' \alpha}$ are defined as
\begin{align}
   \vec G_{\alpha' \alpha} &= \int\, \prod_i d^3\vec r_i \, \psi^*_{\alpha'}(\vec r_i\text{;}\rho_{\alpha'}) \left. \left(\vec \nabla_\rho \psi_\alpha(\vec r_i\text{;}\rho)\right) \right |_{\rho = \rho_\alpha}.
\end{align}
Note that the gradient is evaluated at the equilibrium position $\rho_\alpha$ for the state $\alpha$. Furthermore, only wavefunctions $\psi_{\alpha'}$ with the same symmetry as the ground state contribute to the sum since only those can experience avoided crossings, as opposed to real crossings between states with distinct symmetry. With fixed molecular orientation, the non-adiabatic coupling vectors always point along the molecular axis, so we can write
\be
\vec G_{\alpha' \alpha} \equiv G_{\alpha' \alpha} \, \hat{\rho}_0.
\ee

The wavefunction corrections $\delta \Psi_\alpha$ yield nonzero matrix elements in Eq.~(\ref{eq:NACmatrixEl}), despite the fact that the operator in Eq.~(\ref{eq:NACmatrixEl}) only contains nuclear coordinates, because (for example) $\langle \Psi_\alpha| \delta \Psi_0 \rangle \propto G_{\alpha 0} \langle \psi_\alpha | \psi_\alpha \rangle = G_{\alpha 0}$ by orthonormality of the BO wavefunctions. As we show in Appendix~\ref{app:NAC}, the NAC matrix element for a final state at vibrational level $n$ is given by
\begin{align}
& \mathcal M^{(\alpha)}_{{\rm NAC}, n}
= \frac{i q \eta  G_{\alpha 0}}{\epsilon_\alpha - \epsilon_0}  \nonumber \\
&\times \bigg(\frac{\langle \chi_n^{(\alpha)} |  a_1 e^{  - i \frac{\mu}{M_1} q \rho \eta } | \chi_0 \rangle}{M_1} - \frac{\langle \chi_n^{(\alpha)} |  a_2 e^{  + i \frac{\mu}{M_2} q \rho \eta } | \chi_0 \rangle}{M_2} \bigg) ,
\label{eq:MNAC}
\end{align}
where $\epsilon_0$ is the ground-state electronic energy.

To facilitate comparison to the CMR matrix elements, we can write the NAC probability as
\be
P^{(\alpha)}_{\rm NAC} = \sum_n |\mathcal M^{(\alpha)}_{{\rm NAC}, n}|^2 \equiv P^{(\alpha)}_{e,{\rm NAC}} \times P^{(\alpha)}_{N,{\rm NAC}},
\ee
where
\be
\label{eq:PeNAC}
P^{(\alpha)}_{e,{\rm NAC}} = \frac{q^2 \eta^2 |G_{\alpha 0}|^2}{M^2(\epsilon_\alpha - \epsilon_0)^2}
\ee
and
\begin{align}
\label{eq:PNNAC}
P^{(\alpha)}_{N,{\rm NAC}} = \sum_n &\Bigg( \bigg |\bigg\langle \chi^{(\alpha)}_n \bigg | a_1 \frac{M_2}{\mu} e^{- i \frac{\mu }{M_1} q \rho \eta }  \nonumber \\  
&\;\;\;\;-a_2 \frac{M_1}{\mu} e^{+ i \frac{\mu }{M_2} q \rho \eta } \bigg | \chi_0 \bigg\rangle \bigg |^2\Bigg).
\end{align}
In the homonuclear case ($M_1 = M_2$ and $a_1 = a_2 = a$),
\be
P^{(\alpha)}_{N,{\rm NAC}} \to 16 a^2  \sum_{n} | \langle \chi^{(\alpha)}_n | \sin (q \rho \eta/2)| \chi_0 \rangle|^2 \ \ {\rm (hom.)}.
\ee
The prefactor is larger by a factor of $2^2 = 4$ compared to Eq.~(\ref{eq:PNCMR}), which originates from the fact that the NAC matrix elements scale inversely with the individual nuclear masses rather than the total mass of the molecule. In the case of a larger homonuclear molecule with $N_{n}$ identical atoms, this factor scales as $(M_i\sum_i^{N_n} (M_i)^{-1})^2 = N_n^2$ from reduced mass considerations. Therefore, we might expect that the NAC Migdal effect becomes significantly more dominant for larger molecules.
 
\subsection{Parametric scaling of CMR and NAC}
\label{sec:parametric}

The nuclear matrix elements for CMR and NAC are parametrically identical for diatomic molecules, as can be seen directly from Eqs.~(\ref{eq:PNCMR}) and~(\ref{eq:PNNAC}), up to the factor of 4 mentioned above. Therefore, the parametric scaling of the CMR and NAC components of the Migdal probability $P^{(\alpha)}$ will be determined primarily by the electronic matrix elements. For generic states $\psi_\alpha$ which do not violate selection rules, the dipole matrix element $D_{\alpha 0}$ which governs the CMR rate is proportional to $\azero$, so from Eq.~(\ref{eq:PeCMRWithEta}) we have (dropping factors of the anisotropy parameter $\eta$ for the purposes of this parametric estimate)
\be
P_{e, {\rm CMR}} \sim \left(\frac{m_e}{M}\right)^2 (q \azero)^2,
\ee
as was previously derived for atomic systems~\cite{Baxter:2019pnz,Essig:2019xkx}. For NAC, $\nabla_\rho \sim 1/\azero$ and hence $G_{\alpha 0} \sim 1/\azero$, so we have from Eq.~(\ref{eq:PeNAC}) 
\be
P_{e, {\rm NAC}} \sim \frac{N_n^2 q^2}{M^2 \azero^2 (\Delta E)^2}.
\ee
where we have attached the factor of $N_n^2$ from the nuclear matrix element to emphasize its role for larger molecules. In molecular systems, $\Delta E$ is of order the Rydberg constant $\alpha_{\rm EM}^2 m_e$, and $\azero = (\alpha_{\rm EM} m_e)^{-1}$, where $\alpha_{\rm EM} \simeq 1/137$ is the fine-structure constant. Substituting and rearranging terms yields
\begin{align}
P_{e, {\rm NAC}} & \sim \frac{N_n^2 q^2 \alpha_{\rm EM}^2 m_e^2}{M^2 \alpha_{\rm EM}^4 m_e^2} = N_n^2 \left(\frac{m_e}{M}\right)^2  \left(\frac{q^2}{\alpha_{\rm EM}^2 m_e^2}\right) \nonumber \\
& = N_n^2 \left(\frac{m_e}{M}\right)^2 (q \azero)^2,
\end{align}
which is parametrically identical to $P_{e, {\rm CMR}}$ up to the factor of $N_n^2$.

As we have noted, though, CMR and NAC obey orthogonal selection rules (and thus their scattering amplitudes do not interfere), since the dipole operator $\vec{r}_i$ only connects states of opposite electronic parity while the nuclear gradient $\nabla_\rho$ preserves electronic parity. That said, in molecules where states of both parities have similar energies, we generically expect the CMR and NAC probabilities to be equal within an order of magnitude or so. Note that without including NAC, one might have expected that Migdal transitions which are dipole-forbidden would be suppressed by an additional power of $\left(\frac{m_e}{M}\right)^2 (q \azero)^2 \ll 1$ from expanding the exponential to the next order. In fact, though, the probabilities are much larger; as we will see, NAC typically dominates over CMR in diatomic molecules, due in part to the factor of $N_n^2$.

\subsection{Examples: H$_2^+$, N$_2$, CO}

We calculated the electronic matrix elements relevant for the NAC Migdal effect in N$_2$ and CO using the multi-reference-configuration-interaction (MRCI) method available in MOLPRO 2019.2 \cite{MOLPRO}. The derivative operator
\begin{equation}
\left\langle\psi_{\alpha}(\vec{r_i}; \rho_\alpha) \left|\frac{\partial}{\partial \rho} \right| \psi_{0}(\vec{r_i}; \rho_0)\right\rangle,    
\end{equation}
is numerically implemented as the average of a forward and backward difference scheme using a step size of $0.05 \azero$ around the equilibrium separation $\rho_0$ of the ground electronic state ($2.07 \azero$ for N$_2$~\cite{Herzberg} and $2.13 \azero$ for CO~\cite{database}). For each of the geometries, we employ a Multi-Configuration Self-Consistent Field (MCSCF) calculation with a full valence active space with two frozen orbitals to obtain a set of natural orbitals necessary for the MRCI calculation, in which two states of the same symmetry as the ground state are included. The calculations are carried out by employing the AVQZ basis set \cite{BasisSet} for each atom. As a result, we obtain the matrix elements between the ground electronic state (X$^1\Sigma$) and the first excited state with the same symmetry as
\be
|G_{\alpha 0}|, \ (\epsilon_\alpha - \epsilon_0) = \begin{cases} 
0.64 \azero^{-1}, \ (12.4 \ {\rm eV}), &  {\rm N}_2, \\
1.50 \azero^{-1}, \ (10.8 \ {\rm eV}), & {\rm CO}.
\end{cases}
\label{eq:Galpha0}
\ee
where we have also given the electronic energies of the relevant states with respect to the ground state.

At the same level of theory and basis set, we have also computed the transition dipole moment $D_{\alpha 0}$ between the ground state and the first dipole-allowed electronic state at the equilibrium distance, and find 
\be
|D_{\alpha 0}|, \ (\epsilon_\alpha - \epsilon_0) = \begin{cases} 
0.70 \azero, \ (12.6 \ {\rm eV}), &  {\rm N}_2, \\
0.62 \azero, \ (8.1 \ {\rm eV}), & {\rm CO}.
\end{cases}
\label{eq:Dalpha0}
\ee
In the case of N$_2$, the electronic states $^1\Sigma^{+}_u$ and $^1\Pi_u$ are strongly mixed~\cite{N2_1,N2_2}; however in our case, using two $^1\Sigma^{+}_g$ states and two $^{1}\Pi_u$ states in the MCSCF calculation, we find a transition dipole moment which agrees with the expected range of values due to the strong mixing. In the case of CO, we proceed in the same way. However, since the point group shows C$_{2v}$ symmetry, we include two states of symmetry A$_1$ and two states of symmetries B$_1$ and B$_2$ in the MCSCF calculation, yielding a transition dipole moment which agrees with previous calculations \cite{CO_1,CO_2,CO_3,CO_4}.

In the spirit of treating diatomic molecules as simple toy examples, we also investigated the simplest diatomic molecule, H$_2^+$, which contains a single electron. Indeed, this molecule was studied in the first neutron scattering paper on the Migdal effect~\cite{lovesey1982electron}. Because the 3-body Schr\"{o}dinger equation is separable in the BO approximation, the electronic wavefunctions can be determined by direct numerical integration without needing to approximate them by a basis set of atomic orbitals. We determined the electronic wavefunctions following Ref.~\cite{grivet2002hydrogen}, using a step size of $0.02 \azero$ to calculate the NAC gradients. We find
\be
|G_{\alpha 0}| \ (\epsilon_\alpha - \epsilon_0) = 0.14 \azero^{-1} \ (11.6 \ {\rm eV}), \ \ {\rm H}_2^+.
\ee
However, as we will see in Sec.~\ref{sec:Daily}, the large change in equilibrium separation, from $\rho_0 = 2.04 \azero$ for the ground state to $\rho_\alpha = 8.83 \azero$ for the first NAC state, as well as the large change in the vibrational energies, gives exponentially small overlaps for the nuclear states and hence an atypically small Migdal rate compared to generic diatomic molecules.\footnote{At small $q$, which corresponds to small DM masses, this is equivalent to the statement that the Franck-Condon factor for the transition is very small.} Furthermore, the possible CMR states are so weakly bound that they have only been studied theoretically~\cite{H2plus}, and they have the same issues with large mismatches in the nuclear wavefunctions. As a result, we will focus the subsequent discussion on CO and N$_2$ rather than H$_2^+$.

\subsection{Comparison to inclusive Migdal rates}

In Refs.~\cite{lovesey1982electron,colognesi2005can}, it was noted that for NAC, an approximate sum rule can be used to estimate the inclusive probability $1 - P^{(0)}$ for a transition to any electronic state above the ground state (including the contributions from ionization, rotational nuclear states, and dissociated molecular states):
\be
1 - P^{(0)} \approx \frac{q^2}{M^2 \bar{\epsilon}^2} \langle \nabla_\rho \psi_0 | \nabla_\rho \psi_0 \rangle,
\ee
where $\bar{\epsilon}$ is an ``average'' electronic energy above the ground state, which strictly speaking is ill-defined for an inclusive probability. Unfortunately, $1-P^{(0)}$ can not be calculated with standard quantum chemistry methods since it requires at least two electronic states with the same symmetry. In other words, at least two states are needed to see an avoided crossing associated with the NAC effect. However, we estimate an upper bound on $1 - P^{(0)}$ by considering the inner product of the orbital parts of $\psi_0$ alone. This yields $\langle \nabla_\rho \psi_0 | \nabla_\rho \psi_0 \rangle \simeq 600 \azero^{-2}$ for N$_2$ and $\simeq 700 \azero^{-2}$ for CO. Taking $\bar{\epsilon}$ to be the first ionization potential of the molecule (15.6 eV for N$_2$ and 14.0 eV for CO) as a representative average between bound and continuum states, we can estimate
\be
\frac{P^{(\alpha)}_{{\rm NAC}}}{1-P^{(0)}} \gtrsim \begin{cases} 
1 \times 10^{-3} , &  {\rm N}_2, \\
5 \times 10^{-3}, & {\rm CO}.
\end{cases}
\ee

Ref.~\cite{lovesey1982electron} already calculated $1-P^{(0)}$ for H$_2^+$, finding that it was $10^{4}$ larger than the transition probability $P^{(\alpha)}$ to the first available electronic state, assuming the nuclear wavefunction remains in the $n = 0$ state of the new electronic potential. However, we can understand this large hierarchy between the exclusive and inclusive probabilities as being due to the significant mismatch between the equilibrium separations for the two states. This is to be contrasted with the cases of CO and N$_2$, where the large nuclear overlaps and large values of $G_{\alpha 0}$ result in the transition to the first NAC state giving a contribution to the inclusive rate which is an order of magnitude larger than in H$_2^+$. 

While the above estimates suggest that all of our projected sensitivities in these molecules may be further improved by 2--3 orders of magnitude by using the inclusive excitation rate, as we have alluded to in the Introduction, excitation to continuum states may be more difficult to detect than bound states. Furthermore, the inclusive matrix element is isotropic due to a sum over all possible rotational states, so in order to identify the daily modulation signal, we focus on the exclusive Migdal probabilities as we discuss in the following Sec.~\ref{sec:Daily}.

\section{Daily modulation from the Migdal effect}
\label{sec:Daily}

\subsection{Anisotropies from electronic and nuclear matrix elements}

As we have seen in the previous section, there are two sources of anisotropy in the Migdal probability $P^{(\alpha)}$. The first comes from the dot product of the momentum transfer $\vec{q}$ with either the molecular dipole (for CMR) or the nuclear gradient (for NAC), both of which point along the molecular axis and yield $P^{(\alpha)} \propto \eta^2$. The second comes from the nuclear matrix elements, Eqs.~(\ref{eq:PNCMR}) and~(\ref{eq:PNNAC}), which contain factors of $\eta$ in the exponent. 

\begin{figure}
    \centering
    \vspace{-0.75cm}
    \includegraphics[width=0.48\textwidth]{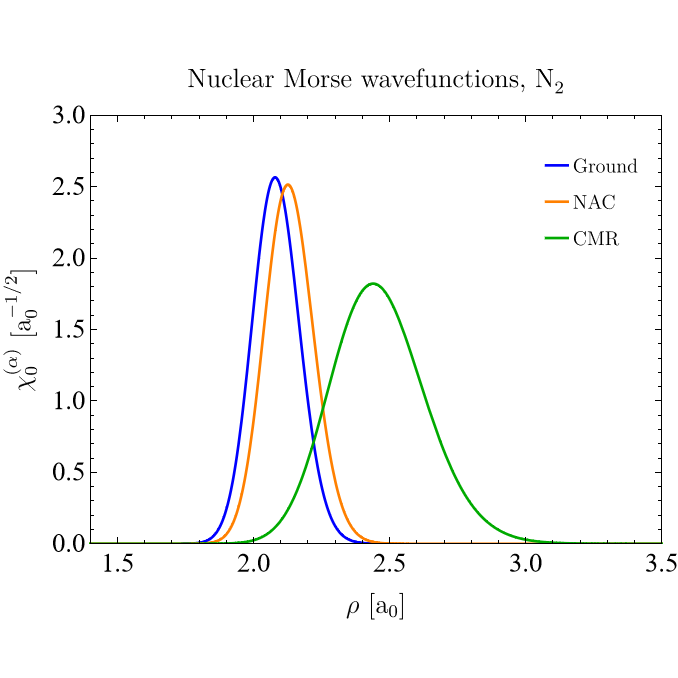}
    \vspace{-1.2cm}
    \caption{Lowest-energy nuclear wavefunctions $\chi_0^{(\alpha)}(\rho)$ for the electronic ground state (blue), first NAC state (orange), and first CMR state (green) in N$_2$. The similarity of the NAC state to the electronic ground state in both the equilibrium separation and the wavefunction spread leads to large nuclear overlaps and a larger rate compared to CMR.}
    \label{fig:GSOverlaps}
\end{figure}

To gain some intuition for the anisotropy from the nuclear matrix elements, consider the case where the states $|\chi^{(\alpha)}_n \rangle$ are governed by the same 1-dimensional harmonic oscillator potential as the ground state $|\chi_0 \rangle$, with the same oscillator frequency $\omega$ and the same equilibrium separation. This is true at the percent level for the NAC transitions in N$_2$ and CO (see Fig.~\ref{fig:GSOverlaps}) due to the fact that these molecules are highly covalent, though not for H$_2^+$ or for the CMR states for N$_2$ and CO. In this case, the matrix elements are~\cite{Kahn:2020fef}:
\be
\label{eq:SHOMatrixElements}
\langle \chi^{(\alpha)}_n | e^{i \widetilde{q}_{1,2} \rho \eta} | \chi_0 \rangle \propto \eta^n \left(\frac{\widetilde{q}_{1,2}}{\sqrt{\mu \omega}}\right)^n \exp \left( - \frac{\widetilde{q}_{1,2}^2 \eta^2}{4 \mu \omega} \right),
\ee
where $\widetilde{q}_{1,2}$ stands for $(\mu/M_1)q$ or $(\mu/M_2)q$ as appropriate and we have dropped normalization constants. Squaring this directly yields a Poisson distribution, but in our case we have to sum over two terms with different $\widetilde{q}$ weighted by $a_{1,2}$. Regardless, it is clear that there is strong dependence on $\eta$ governed by the typical value of the momentum transfer, which is in turn determined by the DM mass. We therefore expect large modulation with amplitude and phase both depending on the DM mass. We derive the general expression for the matrix element, with different oscillator frequencies and equilibrium separations for the initial and final states, in terms of Hermite polynomials in Appendix~\ref{app:Hermite}.

\subsection{Time-dependent rate}

\begin{figure*}[t]
\vspace{-1.0cm}
    \includegraphics[width=0.45 \textwidth]{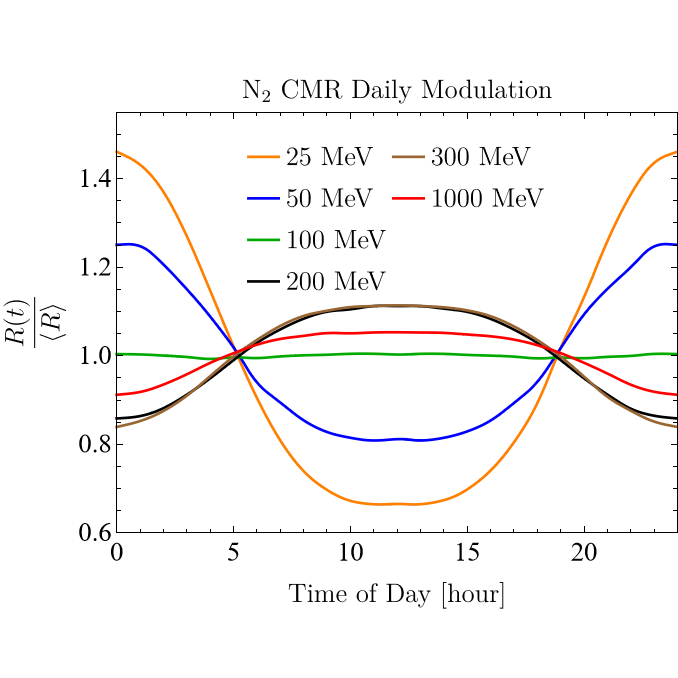} \qquad
    \includegraphics[width=0.45 \textwidth]{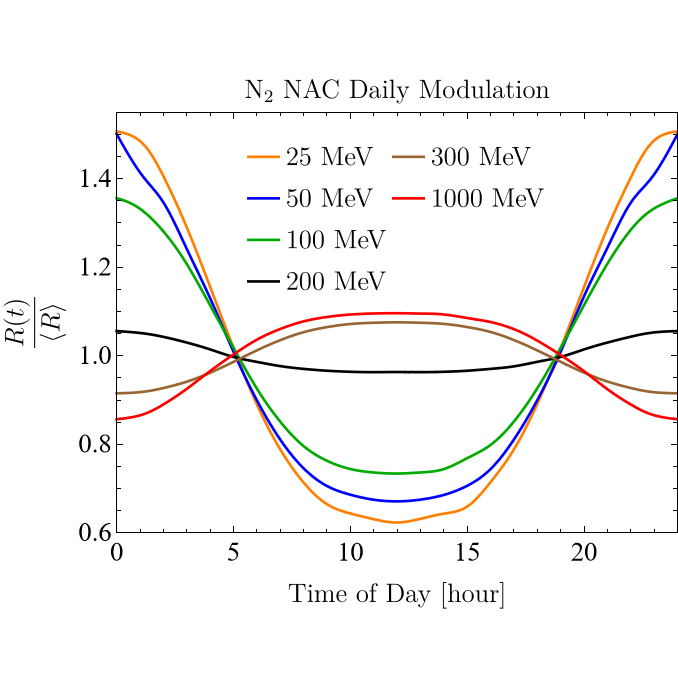}
    \vspace{-1cm}
    \caption{Daily modulation patterns for CMR (left) and NAC (right) in N$_2$. Both components of the molecular Migdal effect exhibit similar behavior, featuring modulation patterns that vary considerably for different DM masses with an inflection point around 200 MeV. The peak-to-trough modulation amplitude saturates to $\simeq 20\%$ at large masses.}
    \label{fig:N2dailymod}
\end{figure*}

\begin{figure*}[t]
\vspace{-0.4cm}
    \includegraphics[width=0.45 \textwidth]{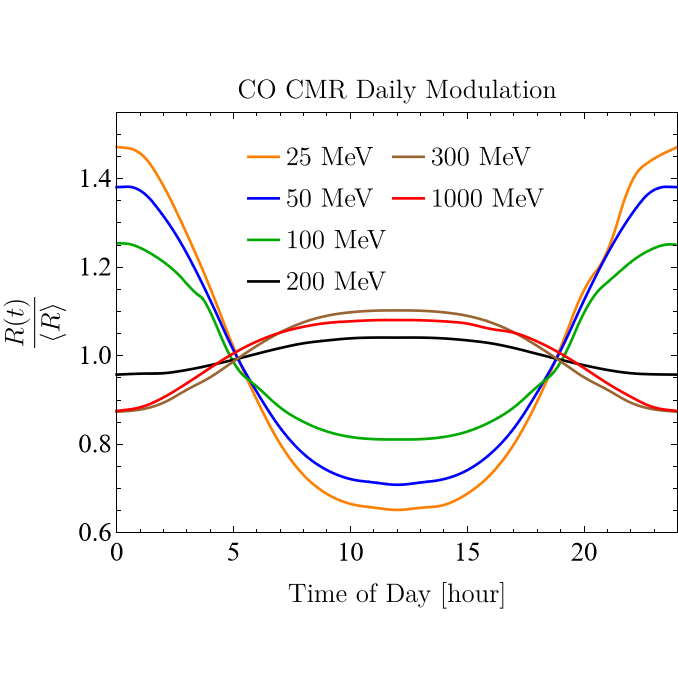} \qquad
    \includegraphics[width=0.45 \textwidth]{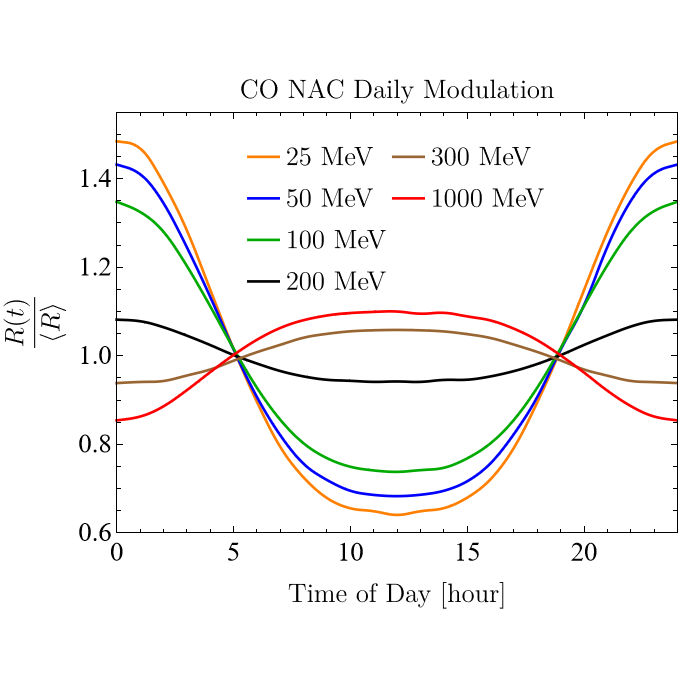}
        \vspace{-1cm}
    \caption{Same as Fig.~\ref{fig:N2dailymod} but for CO. The modulation curves at small and large DM masses are nearly identical to those for N$_2$.}
    \label{fig:COdailymod}
\end{figure*}

For simplicity, consider a model of DM-nuclear scattering where DM couples equally to protons and neutrons. The molecular Migdal rate per unit mass is
\be
    \frac{R^{(\alpha)}(t)}{m_T} = \frac{N_\text{A}}{m_{\rm molar}} \frac{\rho_\chi}{m_\chi} \frac{\bar\sigma_n}{\mu_{\chi n}^2} \int\! \frac{d^3 \vec{q}}{4\pi} g_0(\vec{q},t) F_\text{DM}^2(q) P^{(\alpha)}(\vec{q}),
    \label{eq:RTotg0}
\ee
where $m_T$ is the mass of the target molecule, $m_{\rm molar}$ is its molar mass, $N_A$ is Avogadro's number, $\bar \sigma_n$ is a fiducial DM-nucleon cross section, $\mu_{\chi n}$ is the DM-nucleon reduced mass, $F_\text{DM}(q)$ is the DM form factor which is equal to 1 for a heavy mediator and is proportional to $1/q^2$ for a light mediator. Note that for homonuclear molecules, $P^{(\alpha)} \propto A^2$ where $A$ is the mass number, and we have emphasized that $P^{(\alpha)}$ is a function of the momentum transfer $\vec{q}$ (both magnitude and direction, through the anisotropy parameter $\eta$).

The time dependence of the rate arises from the DM velocity distribution (which we take to be the Standard Halo Model for ease of comparison with the literature) via
\begin{align}
    g_0(\vec{q}, t) = \frac{\pi v_0^2}{q N_0} \left( e^{-v_-(\vec{q},t)^2/v_0^2} - e^{-v_{\rm esc}^2/v_0^2} \right).  \label{vel-g}
\end{align}
Here
\begin{equation} \label{vel-N0}
    N_0 \! = \! \pi^{3/2} v_0^3 \left[ {\rm erf} \left( \! \frac{v_{\rm esc}}{v_0}\! \right) - \frac{2}{\sqrt\pi}\frac{v_{\rm esc}}{v_0} \exp \! \left( \! -\frac{v_{\rm esc}^2}{v_0^2} \! \right)\! \right]
\end{equation}
is a normalization constant depending on the dispersion $v_0 = 220 \ {\rm km/s}$ and the escape velocity $v_{\rm esc} = 544 \ {\rm km/s}$, and
\be
    v_-(\vec{q}, t) = \min \! \left( \! v_{\rm esc}, \frac{\Delta E}{q} + \frac{q}{2m_\chi} + \vec v_\oplus(t) \cdot \hat q \right)
    \label{eq:vMinus}
\ee
is the minimum velocity consistent with energy-momentum conservation, taking $\Delta E$ to be the total energy transfer to the molecule (electronic plus nuclear energies), as shown in Fig.~\ref{fig:TransitionDiagram}. Since $v_-$ arises from integrating the energy-conserving delta function implicit in $P^{(\alpha)}$, as noted in Sec.~\ref{sec:BO} above, it is often sufficient to approximate $\Delta E$ by just the electronic energy, but strictly speaking each term in the sum over the final nuclear states should be weighted by its own $g_0$ with the appropriate value of $\Delta E$ in $v_-$. We adopt conventions consistent with Refs.~\cite{Coskuner:2019odd,Blanco:2021hlm} where the molecular axis $\hat{\rho}_0 = \hat{z}$ points in the direction of the DM wind at $t = 0$, and the Earth velocity is described by
\begin{equation}
\vec{v}_\oplus (t) \! = \! \! |\vec{v}_\oplus| \left( \! \begin{array}{c} \sin\theta_e \sin \vartheta \\ \sin\theta_e \cos\theta_e ( \cos \vartheta - 1) \\ \cos^2 \theta_e + \sin^2 \theta_e \cos\vartheta \end{array} \! \right),
\label{eq:vE}
\end{equation}
where $\vartheta(t) = 2\pi \times \left( \tfrac{t}{24\, \text{h} } \right)$ has the period of a sidereal day, $\theta_e \approx \, 42^\circ$, and we take $|\vec{v}_\oplus| = 234 \ {\rm km/s}$.

\subsection{Daily modulation and sensitivity}

\begin{figure}
    \centering
    \vspace{-1.25cm}
    \includegraphics[width=0.48\textwidth]{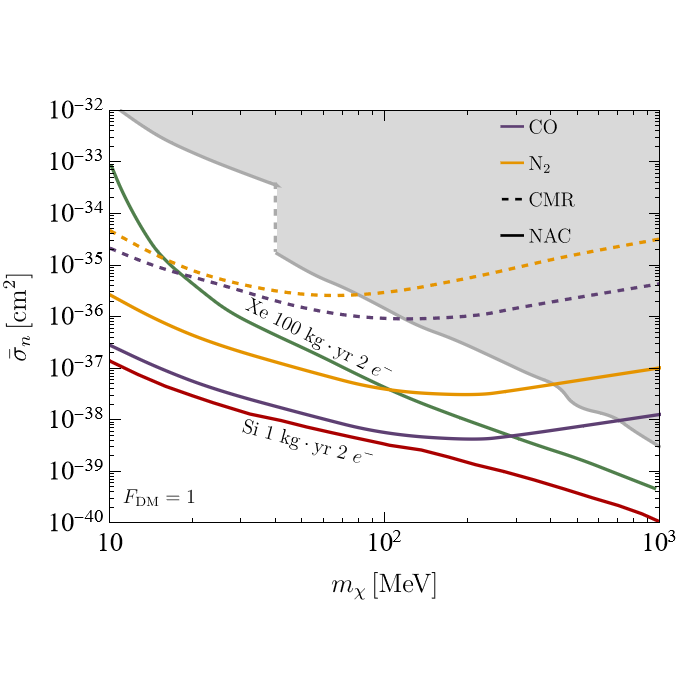}
    \vspace{-1.5cm}
    \caption{3-event background-free projected exclusion limits with 1 kg-yr exposure for the CMR (dashed) and NAC (solid) Migdal effects in CO and N$_2$, assuming 100\% signal efficiency. Current exclusion limits from direct nuclear scattering searches~\cite{CRESST:2019jnq}, dedicated Migdal effect searches~\cite{LUX:2018akb,XENON:2019zpr,CDEX:2019hzn,DarkSide:2022dhx}, and electron recoil searches~\cite{XENON10:2011prx,XENON:2016jmt,XENON:2019gfn} analyzed in terms of the Migdal effect~\cite{Knapen:2020aky} are shown in shaded grey. Projections for xenon~\cite{Essig:2019xkx} (green) and silicon~\cite{Knapen:2020aky,Liang:2022xbu} (red) with a $2e^-$ threshold are shown for comparison; the NAC contribution in CO is competitive with the reach of semiconductors.}
    \label{fig:sensitivity}
\end{figure}

Using Eq.~(\ref{eq:RTotg0}) along with the previously computed $P^{(\alpha)}_{\rm CMR}$ and $P^{(\alpha)}_{\rm NAC}$, we can compute the daily modulation amplitudes $R^{(\alpha)}(t)/\langle R^{(\alpha)} \rangle$ where $\langle R^{(\alpha)} \rangle$ is the time-averaged rate, as well as exclusion limits on $\bar \sigma_n$ for our toy examples. Since $P^{(\alpha)} \propto q^2$, the integrand peaks at large $q$ and heavy-mediator models with $F_{\rm DM} = 1$ will maximize the rate. For H$_2^+$, we numerically determined the complete spectrum of bound nuclear states corresponding to the excited electronic state; however, as alluded to previously, the large difference in equilibrium separations for the two relevant electronic states leads to exponentially small nuclear wavefunction overlaps, and thus we do not consider H$_2^+$ further because it is not a particularly representative example. 

For CO and N$_2$, we modeled the effective potential for the nuclei with a Morse potential. The number of bound nuclear states varies according to the electronic energy and effective potential of the CMR or NAC state: for N$_2$ there are 14 (52) for CMR (NAC), and for CO there are 39 (69) for CMR (NAC). The rapid oscillations of highly-excited nuclear states lead to an oscillatory nuclear matrix element, especially at large $q$, so it is convenient to have a closed-form expression for the matrix element to enable rapid evaluation of the rate. We therefore fit the Morse wavefunctions at level $n$ to harmonic oscillator wavefunctions at the same level $n$ (in order to match the number of nodes) with floating normalization, frequency, and equilibrium separation, and summed up to $n = 10$ using the analytic matrix element derived in Appendix~\ref{app:Hermite} but using the exact Morse potential energy eigenvalues. We validated this calculation by instead approximating the nuclear states as harmonic oscillator states with fixed oscillator frequencies given by the curvature of the Morse potential at the equilibrium separation, finding agreement up to $\mathcal{O}(1)$ factors. As the goal of our calculation is to provide an illustration of the phenomenology of the Migdal effect in molecules, rather than predict a precise sensitivity for a particular experimental implementation, this level of accuracy suffices for our purposes. However, cutting off the sum at $n = 10$ likely underestimates the rate at large DM masses, where highly-excited states dominate; we discuss the modeling uncertainty from nuclear states in Appendix~\ref{app:NuclearModeling}. Accurate modeling of the nuclear states will be important for generalizing our work to larger molecules.

Figs.~\ref{fig:N2dailymod} and~\ref{fig:COdailymod}  illustrate the daily modulation patterns for N$_2$ and CO with $F_{\rm DM} = 1$, for both CMR and NAC. As anticipated, there is a strong dependence on the DM mass, with the rate peaking at $t = 0$ hr for light masses but $t = 12$ hr for heavy masses. The crossover occurs at a mass of about 200--300 MeV independent of the molecular target or the CMR/NAC matrix element; for this DM mass, the argument of the exponent in Eq.~(\ref{eq:SHOMatrixElements}), $q^2/(4 \mu \omega)$, is order-1 for $\omega \sim 0.2 \ {\rm eV}$, $\mu \sim 10 \ {\rm GeV}$, and $q \sim m_\chi v \sim 200-300 \ {\rm keV}$. The large peak-to-trough modulation amplitude -- exceeding a factor of 2 even for DM masses well above the electronic excitation threshold and saturating to $\simeq 20\%$ at large masses -- is comparable to the daily modulation signals in electronic~\cite{Coskuner:2019odd,Geilhufe:2019ndy,Blanco:2021hlm,Hochberg:2021ymx} and phonon~\cite{Griffin:2018bjn,Coskuner:2021qxo} excitation, as well as defect formation~\cite{Budnik:2017sbu,Kadribasic:2017obi,Heikinheimo:2019lwg,Sassi:2021umf}.

Fig.~\ref{fig:sensitivity} shows the projected 3-event background-free exclusion limits on the DM-nucleon cross section $\bar{\sigma}_n$ for N$_2$ and CO, assuming a massive mediator ($F_{\rm DM} = 1$) which couples equally to all nucleons, and a 1 kg-yr exposure. The observable signal would be the photon resulting from the de-excitation of the CMR or NAC state, which from Eqs.~(\ref{eq:Galpha0})--(\ref{eq:Dalpha0}) has energy $\mathcal{O}(10 \ {\rm eV})$, and we assume 100\% photon detection efficiency. For both molecules, the NAC rate (solid) is larger than the CMR rate (dashed), by an order of magnitude for N$_2$ and two orders of magnitude for CO. This is not inconsistent with our arguments in Sec.~\ref{sec:parametric} about the parametric scaling of CMR and NAC, but is simply due to an accumulation of several order-1 factors which all happen to push the rate in the same direction. In particular, the NAC states feature larger nuclear overlaps, or equivalently large Franck-Condon factors, compared to CMR for both molecules, as demonstrated in Fig.~\ref{fig:GSOverlaps}. The fact that NAC dominates is also consistent with previous calculations~\cite{colognesi2005can} which found that NAC was larger than CMR by a factor of $\sim 4$ in neutral H$_2$. Indeed, as discussed in Sec.~\ref{sec:parametric}, the factor of 4 in the nuclear matrix element prefactor suggests that all else being equal, the NAC rate will typically exceed the CMR rate in diatomic molecules, and likely also for larger molecules. The sensitivity begins to decrease around 200 MeV for the same reason the daily modulation crossover occurs at that mass: the exponential suppression in the nuclear matrix elements can only be compensated with highly excited states, which we neglect in the sum because they correspond to molecular dissociation. 

We also show for comparison the existing limits from Migdal searches and direct nuclear recoil searches in noble liquids and solid-state calorimeters, as well as projections for a larger xenon experiment and the Migdal effect in silicon. The sensitivity of diatomic molecules is within a factor of 2 from semiconductors in the mass range 10--100 MeV for the same target mass, which motivates further consideration of more realistic solid-state molecular targets in light of the large daily modulation signal which can further improve the sensitivity in the presence of backgrounds.

\section{Generalizing the molecular Migdal effect to larger molecules}
\label{sec:LargerMolecules}
While the present analysis applies specifically to diatomic molecules, the case of larger molecules is also covered by the general formalism that describes both the CMR and NAC Migdal effects. We relegate the precise generalization to larger molecules and computation of $P^{(\alpha)}$ for experimentally viable molecules to future work, but here we outline the necessary steps. 

The nuclear wavefunctions may be approximated by assuming harmonic oscillator states localized to the equilibrium atomic locations for the relevant electronic states. This captures the essential features of the transition from large molecules to semiconductors, where the Migdal effect may be understood to be mediated by (off-shell) phonons~\cite{Liang:2022xbu}, which are quantized normal mode vibrations. Additionally, rotational excitations are energetically inaccessible in molecular crystals which simplifies the calculation as in the diatomic case.  The electronic amplitudes, however, must be treated more carefully.

The CMR calculation follows from the separation of COM motion from the relative motion of the atoms. In general, the coordinate systems used for larger molecules are more complicated but can be reduced to a COM coordinate and a set of relative coordinates which are relative to either the COM or to the atoms themselves (so-called internal coordinates). Therefore, the computation of the CMR amplitude should proceed identically. The electronic matrix element in Eq.~(\ref{eq:PeCMR}) is related to the oscillator strengths of the electronic transitions, which have been experimentally measured through spectroscopy for most molecular scintillators.

Computing the non-adiabatic coupling vectors $\vec G_{\alpha' \alpha}$ is more difficult as the nuclear gradients become non-trivial with larger and more complicated molecules, which have many more degrees of freedom. In practice this is done through a finite difference method which involves recalculating the electronic molecular orbitals at least six times per atom (three spatial directions for the gradient, evaluated twice for a difference approximation to the gradient). However, the computation simplifies if the nuclear gradients can be computed analytically, for example when the electronic wavefunctions are expressed as linear combinations of atomic orbitals (LCAO)~\cite{abad_calculation_2013}. Such an LCAO approach to molecular orbitals has been shown to be effective in calculating DM-electron scattering rates in organic molecules~\cite{Blanco:2019lrf,Blanco:2021hlm}.

\subsection{Properties of an optimal target}
Using the intuition gained from our simple toy examples, we now turn to an analysis of the physical and chemical properties relevant for maximizing the molecular Migdal effect. We note first that the masses of the atoms in the molecule are not expected to significantly affect the excitation probability, at least for DM coupling equally to all nucleons: $P^{(\alpha)}$ has a factor of $A^2$ in the numerator from coherent scattering from the nucleus, but a factor of $M^2 \propto A^2$ in the denominator, so any coherent enhancement cancels. For the de-excitation photon to be observable, we also need a material which is transparent to its own scintillation light, which could be accomplished by e.g.\ vibrational broadening or lattice effects.

From Eqs.~(\ref{eq:PNCMR}), (\ref{eq:PeCMR}), (\ref{eq:PeNAC}), and (\ref{eq:PNNAC}), we find that there are experimental observables that might indicate that a certain molecule would have a particularly large molecular Migdal amplitude. As mentioned in Sec.~\ref{sec:CMR}, the matrix element in $P^{(\alpha)}_{e,\text{CMR}}$, Eq.~(\ref{eq:PeCMR}), is proportional to the oscillator strength of the electronic transition which can be measured through simple UV-visible absorption experiments. Furthermore, at small $q$, the matrix element in $P^{(\alpha)}_{N,\text{CMR}}$ (\ref{eq:PNCMR}) is equivalent to the Frank-Condon factor for the CMR transition to the state $\alpha$, which can be inferred from the 0-0 substructure of the UV-visible absorption band for this transition. Therefore, in order to determine promising candidates with large CMR molecular Migdal rates, one might look for molecules whose UV spectra show significant, low-lying, dipole-allowed absorption bands which have prominent 0-0 vibrational substructure. 

On the other hand, the matrix element in $P^{(\alpha)}_{e,\text{NAC}}$ (\ref{eq:PeCMR}) is the NAC vector which is a much more subtle molecular object. These non-adiabatic derivative couplings are responsible for the Herzberg-Teller effect, wherein classically forbidden electronic transitions show up in the absorption spectrum of a molecule with pronounced vibrational substructure~\cite{geldof_vibronic_1971,grochala_vibronic_2003,kundu_franckcondon_2022,azumi_what_1977}. Heuristically, this is understood to happen when the forbidden dipole matrix element of the electronic transition depends on the nuclear coordinate which makes the total molecular dipole matrix element non-separable; in the language of Sec.~\ref{sec:BO}, the mismatch between the electronic dipole and the molecular dipole is of order $m_e/M$. In the chemistry literature this is known as ``intensity borrowing'' and is a well-known, experimentally-observed non-adiabatic effect. In fact, the first transition of benzene shows evidence of significant non-adiabatic couplings~\cite{adachi_vibronic_1999,metz_vibronic_1977}. Meanwhile, unlike for CMR, the nuclear matrix element in $P^{(\alpha)}_{N,\text{NAC}}$ vanishes at $q = 0$, so at small $q$ it is dominated by the nuclear dipole of the transition. This vibrational dipole amplitude is non-zero only for integer changes to the vibrational state. Therefore, the 0-1 vibrational substructure of the IR absorption spectra should be proportional to this matrix element. Optimal molecular candidates for the NAC Migdal effect will likely be molecules whose UV-visible spectra show strong, low-lying, dipole-forbidden absorption bands while their IR spectra show significant 0-1 transitions.

We can extend this reasoning to larger molecules, particularly aromatic organic compounds such as benzene and t-stilbene. The vibrational states of these molecules, whose electronic transitions involve delocalized $\pi$-electrons, should not change significantly between electronic states. This is because delocalized $\pi$-electrons involved in the carbon-carbon double bond are not the dominant orbitals which generate the molecular structure, but rather the $\sigma$-electrons on the carbon-carbon single bonds. This chemical structure will generically lead to large Franck-Condon factors and thus large nuclear overlaps, and hence large NAC amplitudes as long as the non-adiabatic couplings are not parametrically small.

\section{Conclusions}
\label{sec:Conclusions}

In this paper we have identified two Migdal effects in molecules wherein DM-nucleus scattering can generate observable electronic transitions. We focus on molecules, rather than isolated atoms or semiconductors, in part because recent work has shown that molecular crystals composed of aromatic molecules feature large anisotropies in their electron-excitation probabilities, leading to $\mathcal{O}(1)$ daily modulation amplitudes in the scintillation signal expected from sub-GeV DM-electron scattering~\cite{Blanco:2021hlm}. The results presented in this paper suggest that these same anisotropies, and therefore the daily modulation, may also be expected in the case of DM-nucleus scattering with an accompanying electronic excitation. We have argued that the sources of anisotropy and the separation of CMR and NAC in diatomic molecules should be qualitatively similar to the case of larger molecules, but with the latter exhibiting a more complex daily modulation pattern due to the richer spectrum of normal modes. We leave the dedicated analysis of larger molecules to future work.

The existence of the NAC and CMR components of the molecular Migdal effect could mean that existing organic scintillators may be used to great effect in constraining the DM-nucleon cross section for masses below $\sim 1$ GeV. Furthermore, we find that CMR is equivalent to the semi-classical Migdal effect, long known for atoms and recently calculated for semiconductors. Our results suggest that the equivalent NAC effect may be present in semiconductors as well, since deviations from the BO approximation are captured by the electron-phonon coupling, though such a calculation (and in particular the relation between CMR and NAC in semiconductors) is beyond the scope of this paper. As discussed above, the NAC Migdal effect in a simple diatomic molecule, carbon monoxide, shows comparable reach per unit mass compared to the projected sensitivity of silicon below about 200 MeV and would outperform xenon in this mass range. Given that diatomic molecules are also poor scintillators with high excitation thresholds, we expect the sensitivities presented here to be a conservative underestimate of the true sensitivities of generic molecular scintillators.

The molecular Migdal effect may also be a promising, though challenging, channel to search for coherent neutrino scattering. The largest flux of solar neutrinos is the low-energy $pp$ spectrum, with an edge at about 400 keV. This yields a maximum nuclear recoil energy of 27 eV for carbon, which is difficult to detect on its own but which can generate an accompanying electronic excitation through the molecular Migdal effect. The coherent neutrino-nucleus scattering rate on carbon is about 1~event/(kg-yr), and the NAC Migdal probability (setting $a = 1$, since the coupling to nucleons is already accounted for in the coherent scattering rate, and $q = 400 \ {\rm keV}$) in CO is about $5 \times 10^{-3}$. In organic crystal detectors, accounting for smaller $\Delta E$ and potentially larger $G_{\alpha 0}$ as well as the ionization and dissociation signals we have neglected, one might optimistically hope to observe a few Migdal events with a 10--100 kg-yr exposure, with some background discrimination possible due to the strong directionality of the signal coming from the Sun. The directionality may also be a useful background discrimination tool for detection of coherent scattering of the much larger flux of keV--MeV reactor neutrinos.

While this study is only a first analysis of a new potential detection channel with molecular detectors, a successful generalization to larger molecules could allow for the reanalysis of existing data (for example, from Ref.~\cite{Blanco:2019lrf}) in order to constrain the DM-nucleus cross section. Furthermore, it opens the possibility for organic scintillator crystals to be used as directional detectors for both DM-electron scattering and DM-nuclear scattering over the entire MeV--GeV mass range. The rich structure of non-adiabatic couplings in molecules is a fruitful area for collaborations between particle physicsists, chemists, and materials scientists, and we look forward to a dedicated exploration of these materials for the next generation of DM detectors.

\noindent \textbf{Acknowledgments.} We thank Duncan Adams, Daniel Baxter, Gordon Baym, John Beacom, Kim Berghaus, Rouven Essig, Danna Freedman, Kathleen Mullin, James Rondinelli, and Lucas Wagner for helpful conversations. C.B. and B.L. are also grateful to the organizers of the Pollica Summer Workshop,  supported by the Regione Campania, Universit\`a degli Studi di Salerno, Universit\`a degli Studi di Napoli ``Federico II'', i dipartimenti di Fisica ``Ettore Pancini'' and ``E R Caianiello'', and ``Istituto Nazionale di Fisica Nucleare'', for hospitality during the completion of this work. The work of C.B.~was supported in part by NASA through the NASA Hubble Fellowship Program grant HST-HF2-51451.001-A awarded by the Space Telescope Science Institute, which is operated by the Association of Universities for Research in Astronomy, Inc., for NASA, under contract NAS5-26555 as well as by the European Research Council under grant 742104. The work of I.H., Y.K., and B.L. was supported in part by DOE grant DE-SC0015655. J. P.-R. acknowledges support from the Simons Foundation.

\appendix

\section{Non-Adiabatic Effects}
\label{app:NAC}
\subsection{Perturbing Hamiltonian}
In this Appendix we derive the effective perturbation, $\delta V$ in Eq.~(\ref{eq:deltaV}) which induces non-adiabatic mixing of electronic states from Eq.~(\ref{eq:deltapsi}). Our derivation follows closely that found in the appendix of Ref.~\cite{lovesey1982electron} and further in Ref.~\cite{MaxBorn1954}. Starting with the Schr\"odinger equation for the molecular energy eigenstates, Eq.~(\ref{eq:SEall}), we plug in the BO ansatz $\Psi = \chi^{(\alpha)}\psi_\alpha$. The nuclear kinetic terms contain
\begin{align}
\label{eq:cross}
\sum_{k=1}^2 \frac{1}{M_k}\nabla_{R_k}^2 \Psi & \supset \sum_{k=1}^2 \frac{2}{M_{k}}(\nabla_{R_k} \chi^{(\alpha)}) \cdot (\nabla_{R_k} \psi_{\alpha} ) \\
& \ll \frac{1}{m_e} \chi^{(\alpha)} \nabla_{i}^2 \psi_{\alpha},
\end{align}
where the inequality follows because $m_e \ll M_k$. Dividing by $\chi^{(\alpha)}$, the cross-term is now small compared to the electronic kinetic energy $\frac{1}{m_e} \nabla^2_i \psi_\alpha$ and can therefore be included as a perturbation to the electronic Schr\"{o}dinger equation (\ref{eq:ElectronicEqSep}). Identifying the operator coefficient of $\psi_\alpha$ as $-2 \, \delta V_1$, we have
\begin{align}
- 2 \, \delta V_1 & = \frac{1}{\chi^{(\alpha)}}\sum_{k = 1}^2 \frac{2}{M_k}  \nabla_{R_k} \chi^{(\alpha)} (\vec{R}_1,\vec{R}_2 ) \cdot \nabla_{R_k}  \\
& = \frac{1}{\chi^{(\alpha)}} \frac{2}{\mu} \nabla_\rho \chi^{(\alpha)} (\vec{\rho}) \cdot \nabla_\rho
\end{align}
where in the second line we have switched to relative coordinates and used $\nabla_{R_2} = - \nabla_{R_1} = \nabla_\rho$ when the COM is fixed and electronic coordinates are neglected; note that the relative minus sign disappears because the gradient is applied twice. We thus identify the perturbing Hamiltonian as
\begin{align}
\delta V_1 &= -\frac{1}{\mu \chi^{(\alpha)}} (\nabla_{\! \rho} \chi^{(\alpha)}_\mu) \cdot \nabla_{\! \rho},
\label{eq:deltaVapp}
\end{align}
which we call $\delta V$ in the main text.

It should be noted that there exists another cross-term in Eq.~(\ref{eq:cross}) which is neglected in the BO approximation given by the following,
\begin{align}
\sum_{k=1}^2 \frac{1}{M_{k}} \chi^{(\alpha)} \nabla_{R_k}^2 \psi_{\alpha}  \ll \frac{1}{m_e} \chi^{(\alpha)} \nabla_{i}^2 \psi_{\alpha}, 
\end{align}
which by similar logic leads to the following electronic perturbing Hamiltonian,
\begin{align}
\delta V_2 &= -\frac{1}{2 \mu} \nabla_{\! \rho}^2 .
\end{align}
However, we will show in the following section that this term is subleading compared to $\delta V_1$ and thus may be neglected in our analysis.

\subsection{Non-Adiabatic Matrix Elements}
\label{sec:AppNAC}
The non-adiabatic coupling comes from the perturbing Hamiltonian $\delta V_1$ in Eq.~(\ref{eq:deltaVapp}). Here we derive the matrix elements resulting from this coupling. We begin by defining convenient rescaled momenta as follows,
\begin{align}
 \vec k_1 &= - \frac{\mu}{M_1} \vec q,
 &
 \vec k_2 &=  \frac{\mu}{M_2} \vec q.
 \label{kay}
\end{align}
Setting aside the scattering lengths $a_i$ for now, we can write the scattering form factors that appear in $\mathcal{M}$,
\begin{align}
\langle \Psi^\prime_\alpha | e^{i \vec k_i \cdot \vec{\rho} } | \Psi_0 \rangle
& = \langle \Psi^{\prime(1)}_\alpha  | e^{i \vec k_i \cdot \vec{\rho} } | \Psi_0^{(0)} \rangle \nonumber \\
& + \langle \Psi^{\prime(0)}_\alpha  | e^{i \vec k_i \cdot \vec{\rho} } | \Psi_0^{(1)} \rangle ,
\end{align}
where 
\begin{align}
\ket{\Psi^{\prime(0)}_\alpha  } &= \ket{\chi_n^{(\alpha)}} \ket{\psi_\alpha} , \\
\ket{\Psi^{\prime(1)}_\alpha } &= \sum_{\alpha'} \frac{ \vec{G}_{\alpha' \alpha} \ket{\psi_{\alpha'} } \ket{ \nabla_\rho \chi_n^{(\alpha)}} }{\mu (\epsilon_{\alpha'} - \epsilon_\alpha) } .
\end{align}
The inner products are given by the following,
\begin{align}
\langle \Psi^\prime_\alpha | e^{i \vec k_i \cdot \vec{\rho} } | \Psi_0 \rangle
&= \sum_{\alpha'}  \frac{\langle \nabla_\rho \chi_n^{(\alpha)}  |\vec G_{\alpha' \alpha}^\star \langle \psi_{\alpha'} | \psi_0 \rangle e^{i \vec k_i \cdot \vec \rho} | \chi_0 \rangle }{\mu (\epsilon_{\alpha' } - \epsilon_\alpha )} \nonumber \\
&+ \sum_{\alpha'}  \frac{\langle  \chi_n^{(\alpha)}  |\vec G_{\alpha' 0} \langle \psi_{\alpha} | \psi_{\alpha'} \rangle e^{i \vec k_i \cdot \vec \rho} | \nabla_\rho \chi_0 \rangle }{\mu (\epsilon_{\alpha' } - \epsilon_0 )} \nonumber
\end{align}
\begin{equation}
= \frac{ \langle \nabla_\rho \chi_n^{(\alpha)} | \vec G_{\alpha0} e^{i \vec k_i \cdot \vec{\rho} } | \chi_0 \rangle  }{\mu (\epsilon_\alpha - \epsilon_0) }
+ \frac{\langle  \chi_n^{(\alpha)} | \vec G_{\alpha0} e^{i \vec k_i \cdot \vec{\rho} } | \nabla_\rho \chi_0 \rangle  }{\mu (\epsilon_\alpha - \epsilon_0) },
\end{equation}
where we used the orthogonality of $\langle \psi_\alpha | \psi_{\alpha'} \rangle = \delta_{\alpha \alpha'} $ as well as the antisymmetry of the coupling vectors, $\vec G^\star_{\alpha'\alpha}=-\vec G_{\alpha\alpha'}$.

The inner product becomes
\begin{align}
\langle \Psi^\prime_\alpha | e^{i \vec k_i \cdot \vec{\rho} } | \Psi_0 \rangle &= \frac{1}{\mu (\epsilon_\alpha  - \epsilon_0) }
\bigg(\langle \nabla_\rho \chi_n^{(\alpha)} |\vec  G_{\alpha0} e^{i \vec k_i \cdot \vec \rho } | \chi_0 \rangle \nonumber \\
&+ \langle  \chi_n^{(\alpha)} |\vec G_{\alpha0} e^{i \vec k_i \cdot \vec \rho } |\nabla_\rho \chi_0 \rangle
\bigg) \nonumber
\end{align}
\begin{align}
& =- \frac{1}{\mu (\epsilon_\alpha - \epsilon_0) }  \vec G_{\alpha0} \cdot  \langle \chi_n^{(\alpha)} | i \vec k_i e^{ i \vec k_i \cdot \vec{\rho} } | \chi_0 \rangle,
\label{eq:NACinnProd}
\end{align}
where we have integrated by parts in the last line and taken the surface term to be zero due to normalizability. Furthermore, note that $\vec G_{\alpha 0}$ factors out of the inner product since it is constant in $\rho$ to first order.

Combining Eq.~(\ref{eq:NACinnProd}) with Eq.~(\ref{eq:NACmatrixEl}) and replacing $\vec{k}_i$ with the original expressions in terms of $M_i$ and $\vec q$:
\begin{align}
\mathcal{M} &= \frac{ i}{M_1 (\epsilon_\alpha - \epsilon_0) } \vec G_{\alpha 0} \cdot  \langle \chi_n^{(\alpha)} |  \vec q e^{  - i \frac{\mu}{M_1} \vec q \cdot \vec \rho } | \chi_0 \rangle \nonumber \\
&- \frac{i }{M_2 (\epsilon_\alpha - \epsilon_0 ) } \vec G_{\alpha 0} \cdot  \langle \chi_n^{(\alpha)} |  \vec q e^{  + i \frac{\mu}{M_2} \vec q \cdot \vec \rho } | \chi_0 \rangle \nonumber \\
&= \frac{i  \vec G_{\alpha 0}\cdot \vec{q}  }{\epsilon_\alpha - \epsilon_0}  \nonumber \\
&\times \bigg(\frac{\langle \chi_n^{(\alpha)} |  a_1 e^{  - i \frac{\mu}{M_1} \vec q \cdot \vec \rho } | \chi_0 \rangle}{M_1} - \frac{\langle \chi_n^{(\alpha)} |  a_2 e^{  + i \frac{\mu}{M_2} \vec q \cdot \vec \rho } | \chi_0 \rangle}{M_2} \bigg), 
\end{align}
which matches the results of Ref.~\cite{lovesey1982electron} in the case of homonuclear molecules.

Finally, recall that the other neglected cross term proportional to $\chi \nabla^2_{R_k}\psi$ generated a perturbing Hamiltonian $\delta V_2 = - \nabla_{\! \rho}^2 /(2\mu)$. Following a similar derivation as above, one can show that this Hamiltonian leads to a matrix element which is proportional to the following factor,
\begin{align}
    \mathcal{M} \sim \langle \chi_n^{(\alpha)} |  a_i e^{  + i \vec k_i \cdot \vec \rho } (\vec \nabla_{\rho} \cdot \vec G_{\alpha 0})  | \chi_0 \rangle.
\end{align}
However, since $\vec{G}_{\alpha 0}$ is independent of $\vec{\rho}$ to first order, this matrix element is a subleading non-adiabatic coupling which we can take to be zero at this order in the expansion.

\section{Harmonic Oscillator Matrix Elements}
\label{app:Hermite}

In this Appendix we derive a closed-form analytic expression for the matrix element $\langle \chi_n^{(\alpha)} | e^{i \beta q \rho} | \chi_0 \rangle$, which appears in the CMR and NAC nuclear matrix elements. Here $\beta$ is an arbitrary real parameter, and the initial and final states are 1-dimensional harmonic oscillator states:
\begin{align}
    \chi_0(\rho) &= \left( \frac{ \mu \omega_0 }{\pi} \right)^{1/4} e^{ -\frac{\mu \omega_0 (\rho - \rho_0)^2}{2}}, \\ 
    \chi_n^{(\alpha)}(\rho) &= \frac{1 }{\sqrt {2^n n!} } \left( \frac{ \mu \omega_\alpha }{\pi } \right)^{1/4} e^{ - \frac{\mu \omega_\alpha (\rho - \rho_\alpha)^2}{2}} \nonumber \\
    & \times H_n\! \left( \sqrt{ \mu \omega_\alpha } (\rho - \rho_\alpha) \right),
\end{align}
with $\mu$ the reduced nuclear mass, $\rho_0$ and $\omega_0$ the equilibrium separation and oscillator frequency for the ground state (likewise for $\rho_\alpha$ and $\omega_\alpha$ for the excited electronic state) and $H_n$ the Hermite polynomials.

It is most convenient to work with momentum-space wavefunctions,
\begin{align}
    \tilde{\chi}_0(k) &= \frac{e^{-i k \rho_0} }{ (\mu \omega_0 \pi)^{1/4} }   \exp\left( - \frac{1}{2} \frac{k^2}{\mu \omega_0} \right)  ,
    \\
    \tilde{\chi}_n^{(\alpha)}(k) &= \frac{e^{-i k \rho_\alpha} }{\sqrt{2^n n!} } \left( \frac{1}{\mu \omega_\alpha \pi } \right)^{1/4} (-i)^n \exp\left( - \frac{1}{2} \frac{k^2}{\mu \omega_\alpha} \right) \nonumber \\
    & \times H_n\!\left( k / \sqrt{ \mu \omega_\alpha} \right),
\end{align}
which also offer the advantage that the translations in $\rho$ appear as overall phase factors. The normalization is chosen so that factors of $1/\sqrt{2\pi}$ appear in both the Fourier transform and its inverse:
\begin{align}
    \tilde{\chi}(k) &= \int\!\frac{d\rho}{\sqrt{2\pi}} e^{-i k \rho } \chi(\rho) ,
    &
    \chi(\rho) &= \int\!\frac{dk}{\sqrt{2\pi}} e^{+i k \rho } \tilde{\chi}(k).
\end{align}
Computing the matrix element in Fourier space yields
\begin{align}
    & \langle \chi_n^{(\alpha)} | e^{i \beta q \rho} | \chi_0 \rangle 
    = \int \frac{d\rho}{2\pi} dk_1 dk_2 \, e^{-i k_1 \rho } [\tilde\chi_n^{(\alpha)}(k_1)]^\star \nonumber \\ & \qquad \qquad \qquad \qquad \times e^{i \beta q \rho} e^{i k_2 \rho} \tilde\chi_0(k_2) \nonumber \\
    &= \int\! dk_1 dk_2 \, \delta(k_1 - \beta q - k_2 ) \, [\tilde\chi_n^{(\alpha)} (k_1) ]^\star \tilde\chi_0(k_2)\nonumber \\
    &= \int_{-\infty}^\infty \! dk \, [\tilde\chi_n^{(\alpha)}(k + \beta q)]^* \tilde\chi_0(k) \nonumber
    \\
    &= \frac{i^n \, e^{i \rho_0 \beta q} }{(\mu^2 \omega_\alpha \omega_0)^{1/4} \sqrt{ 2^n n! \pi} } \nonumber \\
    & \times \int\! dk \, e^{i k (\rho_\alpha - \rho_0) } \exp\!\left( - \frac{1}{2} \frac{k^2 }{\mu \omega_\alpha}  - \frac{1}{2} \frac{(k-\beta q)^2 }{\mu \omega_0} \right) \nonumber \\
    & \ \ \times H_n\! \left( \frac{k }{\sqrt{ \mu \omega_\alpha } } \right) .
\end{align}
One can rearrange the integrand by completing the square. Define
\begin{align}
k_0 & \equiv \sqrt{\frac{2 \mu \omega_0 \omega_\alpha}{\omega_0 + \omega_\alpha}}, \\
\delta \rho &\equiv \rho_\alpha - \rho_0, \\
t_\beta &\equiv \frac{k_0 \beta q }{2 \mu \omega_0 } + i \frac{k_0 \delta \rho }{2},
\end{align}
where the notation $t_\beta$ emphasizes that this variable carries the dependence on the parameter $\beta$. 
The matrix element becomes
\begin{align}
& \langle \chi_n^{(\alpha)} | e^{i \beta q \rho} | \chi_0 \rangle 
 = \frac{i^n \, e^{i \rho_0 \beta q} e^{t_\beta^2 - \beta^2 q^2 / (2 \mu \omega_0) } k_0 }{(\mu^2 \omega_\alpha \omega_0)^{1/4} \sqrt{ 2^n n! \pi} } \nonumber \\
& \times 
\int\! \frac{dk}{k_0} \, e^{- k^2/k_0^2 + 2 (k/k_0) t_\beta - t_\beta^2 } H_n(k/\sqrt{\mu \omega_\alpha} ) .
\end{align}

We will now manipulate the integral to obtain the Hermite generating function:
\begin{align}
e^{2xt - t^2 } &= \sum_{j=0}^\infty \frac{t^j}{j!} H_j(x). 
\end{align}
Taking $x = k/k_0$ and 
\begin{align}
u \equiv \frac{k_0}{\sqrt{\mu \omega_\alpha}}  = \sqrt{ \frac{2 \omega_0}{\omega_0 + \omega_\alpha} } ,
\end{align}
the desired integral is
\begin{align}
\mathcal I &\equiv \int\! dx \, e^{- x^2 + 2 x t_\beta - t_\beta^2 } H_n( uk ) \nonumber \\
& = \sum_{k=0}^\infty \frac{t_\beta^k}{k!} \int\! dx \, e^{-x^2}\, H_k(x)   H_n( uk ).
\label{eq:IHH}
\end{align}
Combining the series definition of the Hermite polynomials and its inverse,
\begin{align}
\label{eq:HermiteSeries}
H_n(a x) &= n! \sum_{m=0}^{\lfloor n/2 \rfloor } \! \frac{(-1)^m (2a x)^{n - 2m} }{m! (n - 2m)!}, \\
(2 x)^s &= s! \sum_{j=0}^{\lfloor s/2 \rfloor} \! \frac{H_{s - 2j}(x) }{j! (s - 2j )! },
\end{align}
yields a series expansion for $H_n(ax)$ in terms of $H_n(x)$:
\begin{align}
H_n(a x) &= n! \sum_{m=0}^{\lfloor n/2 \rfloor } \! \frac{(-1)^m a^{n - 2m} }{m! } \sum_{j=0}^{\lfloor n/2 - m \rfloor } \frac{H_{n-2m - 2j } (x) }{j! ( n - 2m - 2j)!} \nonumber \\
&= \sum_{j=0}^{\lfloor n/2 \rfloor} a^{n- 2j} (a^2 - 1)^j \frac{n!}{(n - 2j)! j!  } H_{n - 2j}(x).
\end{align}
This permits Eq.~(\ref{eq:IHH}) to be integrated term by term using the orthogonality relation
\be
\int dx \, e^{-x^2} H_m(x) H_n(x) = \sqrt{\pi} \, 2^n \, n! \, \delta_{nm},
\ee
which gives
\begin{align}
\mathcal{I} & = \sum_{k = 0}^\infty \sum_{j = 0}^{\lfloor n/2 \rfloor} \frac{t_\beta^k}{k!} \sqrt{\pi} 2^k k! \, \delta_{k, n-2j} \frac{u^{n-2j } (u^2 - 1)^j n! }{(n- 2j)! j!} \nonumber
\\
& =  n! \sqrt{\pi } \sum_{j = 0}^{\lfloor n/2 \rfloor}  \frac{ (u^2 - 1)^j  }{(n- 2j)! j!} (2 u t_\beta)^{n - 2j}.
\end{align}
The argument of the sum can now be manipulated to yield the Hermite series definition, Eq.~(\ref{eq:HermiteSeries}). In fact,
\begin{align}
\mathcal I &= \sqrt{\pi} \left( \sqrt{ 1 - u^2 } \right)^n \, n! \sum_{j=0}^{\lfloor n/2 \rfloor} \frac{(-1)^j}{j! (n-2j)!} \left( \frac{ 2 u t_\beta }{\sqrt{ 1 - u^2 } } \right)^{n - 2j} \nonumber
\\
&= \sqrt{\pi} (1 - u^2)^{n/2} \,H_n\!\left( \frac{u t_\beta}{\sqrt{ 1 - u^2} } \right).
\end{align}
The argument of $H_n$ is generically complex if $\delta\rho \neq 0$ because $t_\beta$ is complex, so the Hermite polynomials are to be understood in their analytic continuation. Restoring the prefactor, we have finally
\begin{align}
\langle \chi_n^{(\alpha)} | e^{i \beta q \rho} | \chi_0 \rangle 
& = \frac{i^n \, e^{i \rho_0 \beta q} e^{t_\beta^2 - \beta^2 q^2 / (2 \mu \omega_0) } k_0 }{(\mu^2 \omega_\alpha \omega_0)^{1/4} \sqrt{ 2^n n!} }  \nonumber \\
& \times (1 - u^2)^{n/2} \,H_n\!\left( \frac{u t_\beta}{\sqrt{ 1 - u^2} } \right).
\label{eq:SHOAnalyticMagic}
\end{align}
This closed-form expression permits rapid evaluation of the required nuclear matrix elements even up to large values of $n$.

\section{Modeling the Nuclear Wavefunctions}
\label{app:NuclearModeling}

\begin{figure*}[t]
\vspace{-1.1cm}
    \includegraphics[width=0.45 \textwidth]{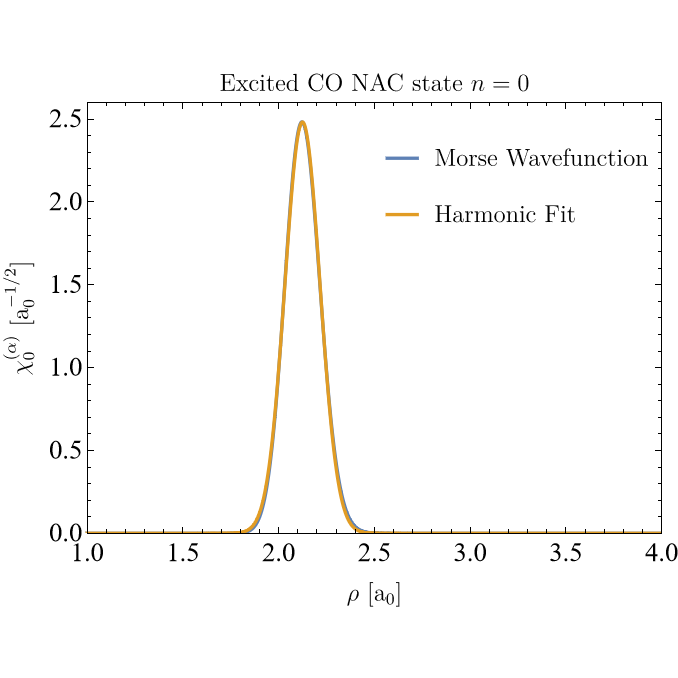} \qquad
    \includegraphics[width=0.45 \textwidth]{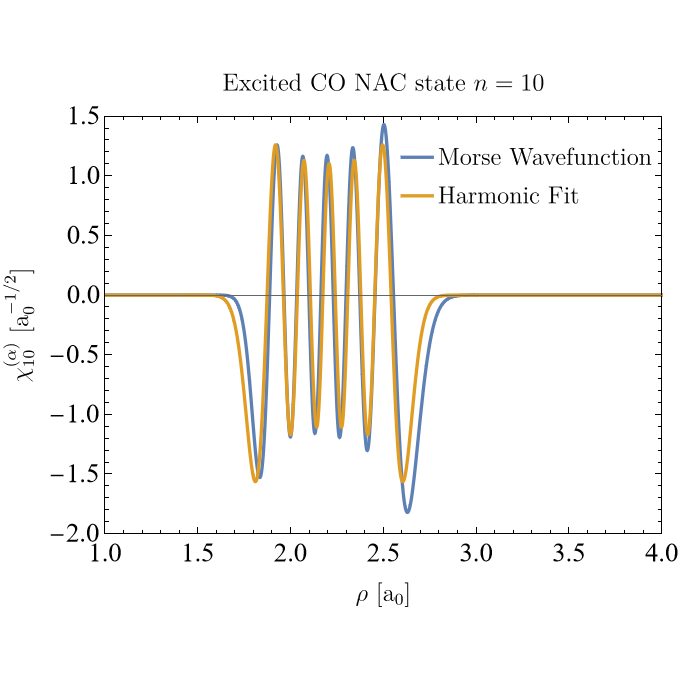}
    \vspace{-1cm}
    \caption{Fits to Morse wavefunctions with harmonic oscillator wavefunctions at the same level $n$. The fit is excellent for the ground state but becomes progressively worse for higher excited states which probe the anharmonicity of the Morse potential.}
    \label{fig:MorseFit}
\end{figure*}

\begin{figure*}[t]
\vspace{-0.5cm}
    \includegraphics[width=0.45 \textwidth]{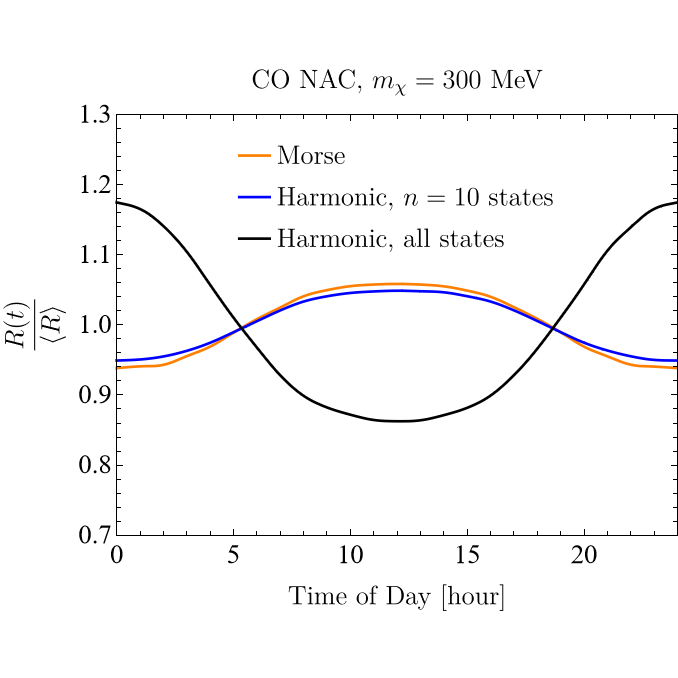} \qquad
    \includegraphics[width=0.45 \textwidth]{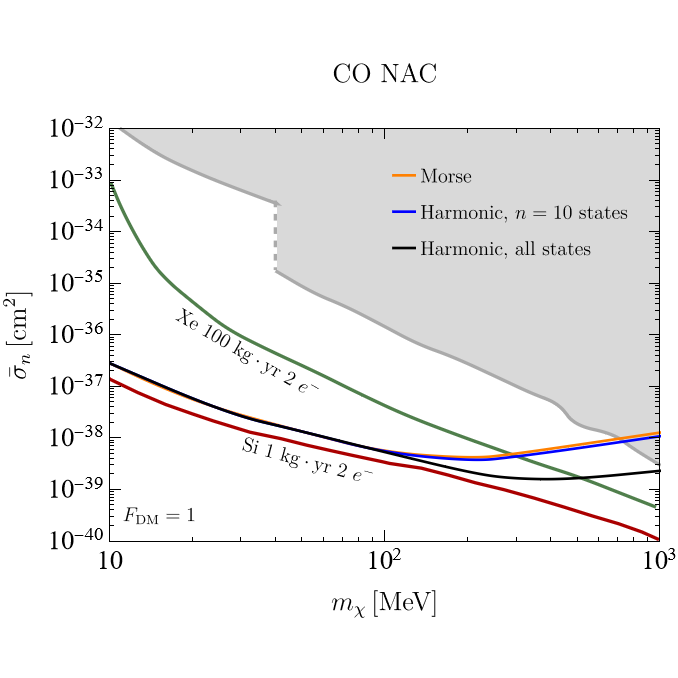}
    \vspace{-1cm}
    \caption{Effect of changing the nuclear wavefunction sum on the daily modulation curves for $m_\chi = 300 \ {\rm MeV}$ (left) and the 3-event sensitivity (right). The choice of wavefunction has a relatively small effect compared to the much larger effect of the cutoff level for the nuclear states. The difference in the phase of the modulation is especially pronounced near the crossover mass of 300 MeV, while the sensitivity is primarily affected at masses above this crossover.}
    \label{fig:MorseCompare}
    \vspace{-0.2cm}
\end{figure*}

In the main body, we fit the Morse potential wavefunctions to harmonic oscillator wavefunctions at the same level $n$ in order to exploit the analytic formula~(\ref{eq:SHOAnalyticMagic}). Fig.~\ref{fig:MorseFit} shows the results of such fits for the CO NAC states; the $n = 0$ ground state (left) is extremely well approximated by the harmonic oscillator ground state, but a highly excited state ($n = 10$, right) has a poorer fit at both large and small $\rho$ where the anharmonicity is most pronounced.

As shown in Fig.~\ref{fig:MorseCompare}, the choice of wavefunction matters only at the percent level for NAC, while the choice of where to cut off the sum in Eqs.~(\ref{eq:PNCMR}) and~(\ref{eq:PNNAC}) is a much larger effect. In Fig.~\ref{fig:MorseCompare}, the black curves labeled ``Harmonic, all states'' extend the sum to the largest value of $n$ such that the nuclear energy $(n+ \frac{1}{2})\omega_\alpha - \frac{1}{2}\omega_0$ does not exceed the depth of the Morse potential (8.96 eV for the first NAC state in CO), yielding $n = 34$. At the crossover mass of 300 MeV where $q^2/(4 \mu \omega_\alpha)$ is $\mathcal{O}(1)$, the extra states in the sum cause the daily modulation to switch phases, from a maximum at $t = 0$ hr to a maximum at $t = 12$ hr. Likewise, for masses above the crossover mass where many highly-excited states contribute to $P^{(\alpha)}_N$, the sensitivity improves by almost a factor of 10. We conclude that our sensitivity estimates using the Morse spectrum cut off at $n = 10$ are robust at the order of magnitude level, but a precise prediction for the molecular Migdal rate will require accurate modeling of the highly-excited nuclear states of the molecular target. 

\bibliography{main}

\end{document}